\documentclass[journal,10pt]{IEEEtran}

\usepackage[super]{nth}
\usepackage{enumitem}
\usepackage{makecell}

\usepackage[T1]{fontenc}
\usepackage{xcolor}
\usepackage{fancyhdr}
\usepackage{tgschola}
\usepackage{lastpage}
\usepackage[caption=false]{subfig}
\usepackage{amsthm,amssymb,amsmath,mathtools,fixmath}
\usepackage{amsbsy}
\usepackage{dsfont}
\usepackage{moreverb}
\usepackage{amsfonts}
\usepackage{graphicx}
\graphicspath{{SimulationFigures/}}
\usepackage{tikz}
\usepackage{listings}
\usepackage{color}
\usepackage{latexsym}
\usepackage{animate}
\colorlet{linkequation}{blue}
\usepackage[linesnumbered,ruled]{algorithm2e}
\usepackage{soul}
\usepackage[T1]{fontenc}
\usepackage[utf8]{inputenc}
\usepackage{tabularx,ragged2e,booktabs}
\usepackage{tabu}
\DeclareMathOperator*{\argminA}{arg\,min} 
\DeclareMathOperator*{\argmaxA}{arg\,max} 

\SetKwInput{KwInput}{Input}                   
\SetKwInput{KwOutput}{Output}              
\SetKwInput{KwInitial}{Initialization}         
\SetKwFor{Loop}{Loop}{}{EndLoop}

\usepackage{lipsum}

\DeclareMathOperator*{\maxi}{maximize}
\makeatletter
\def\BState{\State\hskip-\ALG@thistlm}
\makeatother

\usepackage{geometry}
\usepackage{cite}

\ifCLASSINFOpdf
\else
\fi

\usepackage{amsmath}
\usepackage[cmintegrals]{newtxmath}

\begin{document}

\title{\huge Multi-Agent Q-Learning for Real-Time Load Balancing User Association and Handover in Mobile Networks}

\author{\normalsize Alireza~Alizadeh~\IEEEmembership{\normalsize Member,~IEEE}, Byungju Lim~\IEEEmembership{\normalsize Member,~IEEE}, and
        \normalsize Mai~Vu,~\IEEEmembership{\normalsize Senior~Member,~IEEE}
        \thanks{This work was supported in part by the National Science Foundation under the CNS/SWIFT Grant 2127648.}
\thanks{A. Alizadeh and M. Vu are with the Department of Electrical and Computer Engineering, Tufts University, Medford, 02155, USA (E-mail: $<$alireza.alizadeh,~mai.vu$>$@tufts.edu)}
\thanks{B. Lim is with the Department of Electronic Engineering, Pukyong National University, Busan, 48513, South Korea (E-mail: limbj@pknu.ac.kr)}}



\maketitle

\begin{abstract}
As next generation cellular networks become denser, associating users with the optimal base stations at each time while ensuring no base station is overloaded becomes critical for achieving stable and high network performance. 
We propose multi-agent online Q-learning (QL) algorithms for performing real-time load balancing user association and handover in dense cellular networks. The load balancing constraints at all base stations couple the actions of user agents, and we propose two multi-agent action selection policies, one centralized and one distributed, to satisfy load balancing at every learning step. 
In the centralized policy, the actions of UEs are determined by a central load balancer (CLB) running an algorithm based on swapping the worst connection to maximize the total learning reward. In the distributed policy, each UE takes an action based on its local information by participating in a distributed matching game with the BSs to maximize the local reward. 
We then integrate these action selection policies into an online QL algorithm that adapts in real-time to network dynamics including channel variations and user mobility, using a reward function that considers a handover cost to reduce handover frequency. 
The proposed multi-agent QL algorithm features low-complexity and fast convergence, outperforming 3GPP max-SINR association. Both policies adapt well to network dynamics at various UE speed profiles from walking, running, to biking and suburban driving, illustrating their robustness and real-time adaptability.
\end{abstract}


\IEEEpeerreviewmaketitle

\section{Introduction} \label{Intro}

The ever-increasing demand for higher data rates and lower latencies in beyond 5\textsuperscript{th} generation (B5G) and the upcoming 6th generation (6G) systems pushes for the deployment of dense millimeter wave (mmWave)-enabled cellular networks. These networks have a multi-tier structure composed of multiple base stations (BSs) equipped with different antenna arrays, transmitting at different power levels, operating at different frequency bands from sub-6 GHz to mmWave, and providing different amount of bandwidths. 
In such dense and heterogeneous networks, network control and resource allocation problems become more complicated and require more efficient approaches. 

These dense networks with a high density of BSs intrinsically require user association: the process of assigning each user equipment (UE) to a BS to maximize a network utility. A standing challenge for user association is to satisfy load balancing, that is to keep the number of UEs connected to each BS below its quota (the number of UEs each BS can serve simultaneously), assuming each UE can only connect to one BS at any time instant. 
If load balancing fails, some BSs may suffer from heavy load condition when many users are connected to that BS. In such cases, user connections could be dropped, or the quality of service of associated users can be significantly degraded due to overloaded traffic at that BS and consequent bottleneck on the backhaul link.
Avoiding this scenario necessitates load balancing user association.

This problem, however, is known to be NP-hard due to the presence of integer association variables \cite{Andrews}. The complexity further increases when user association needs to be performed in a highly dynamic network, leading to frequent handovers as a serious concern. As such, there is a need for effective handover algorithms that can adapt to network dynamics including channel variations and user mobility, while maintaining load balancing and achieving high network throughput at a low handover rate.

\subsection{Related and Prior Works}
Various analytical approaches have been proposed to address load balancing user association by relaxing the unique association constrains then solving the relaxed problem to obtain association results \cite{Andrews,Caire,TVT_irs}.
In an earlier work, we introduced a new interference-dependent user association formulation by considering the dependency between user association and interference in mmWave networks \cite{TWC}, which achieves near-optimal solution and has been used as a benchmark in other works including \cite{LimitedCSI20,Khosravi20}.
These algorithms are centralized, hence often exhibiting a high computational complexity and requiring a significant signaling overhead to collect information at a central server. 
In addition, these works studied the user association problem under a static setting without considering user mobility. In a mobile network, however, the effects of abrupt mmWave channel variations and user mobility on user association and handover can be significant. 

In recent years, machine learning has attracted attention of researchers in wireless communications. Reinforcement learning is a method which can take into account real-time interactions with the system without requiring extensive training, making it particularly suitable for a dynamic wireless system \cite{ML_Intro_Book}. 
QL is a specific type of reinforcement learning, which has been applied for user association in the literature, albeit without specific load balancing for all BSs. For example, in dense and static networks, a multi-agent QL model performs for joint power optimization and user association in \cite{QL_UA19}. Deep QL algorithms are used for user association in LTE network \cite{DRL_UA19}, in IoT network \cite{DRL_assoc1}, and in dense mmWave networks \cite{Sana19}. A centralized handover algorithm using QL achieves link-beam performance gain in \cite{5GHandover}.
Handover management can use double deep Q-networks  for vehicle-to-network communications under high mobility vehicles \cite{new_ref4} and for 5G mmwave network \cite{new_ref3}.
These works did not specify a load constraint for every individual BS, often SBSs can offload to the MBS which does not have a specified load limit. 

Recently, several works have considered load balancing using reinforcement learning. 
For example, a decentralized user association algorithm utilizes a multi-agent actor-critic network in a dense mmWave network, considering handover to reduce energy consumption and delay \cite{new_ref1}.
A deep reinforcement learning framework is designed to learn handover parameters and antenna tilt angle for distributing cell load evenly where the load is defined as the ratio of allocated resource blocks (RBs) versus the total number of RBs \cite{new_ref2}.
These works, however, cannot strictly guarantee the load balancing constraints, since these constraints are only considered implicitly and may be violated during online implementation.
Other recent works have used multi-armed bandit (MAB) based centralized and semi-distributed algorithms \cite{TWC_alireza} or distributed deep Q-network (DQN) models \cite{wcl_lim} to strictly satisfy the load balancing constraint for practical implementation.
However, MAB based algorithms in \cite{TWC_alireza} output an action without any information about the network state, hence they may not scale well nor provide the best adaptive action under channel variations and user mobility. The DQN models in \cite{wcl_lim}, while adapting to network states, do not consider mobility nor a handover cost, which can result in high handover rates and significant overhead in a highly dynamic network.




\subsection{Our Contributions}

We propose online multi-agent QL algorithms for user association and handover in dense and highly-dynamic mmWave networks.  
Our proposed algorithms explicitly account for the quota at each BS to ensure that the resulting association always satisfy load balancing for every BS at every learning step. 
The load balancing constraints at BSs introduce dependency among the agents as their actions are coupled, and this dependency is explicitly handled in our proposed algorithms in both centralized and distributed versions. 
To the best of our knowledge, this is the first work proposing 
multi-agent QL models for user association and handover that ensure load balancing at all BSs across all tiers, while effectively adapting in real-time to user mobility and channel variations.

Specifically, we formulate the load balancing user association and handover problem as a multi-agent QL system with UEs as agents, by  defining the states, actions, and update rule of the Q-values. We use the upper confidence bound (UCB) as the metric for action selection.
\textcolor{black}{While our system implements the tabular QL method at each agent, the proposed multi-agent policies and online QL framework also work for deep QL methods. Our main contributions can be summarized as follows. 
}

\begin{itemize}[leftmargin=*]
\item
\textcolor{black}{
We propose two multi-agent action selection policies, a centralized and a distributed version. Both policies ensure the satisfaction of load balancing at all BSs at every learning step. }
The centralized action selection policy is performed at a central load balancer (CLB) to maximize the total learning rewards. The distributed policy is implemented via a matching game to obtain load-balanced actions. 
\textcolor{black}{Both policies require no information exchange among agents and have low signaling overheads.}

\item 
\textcolor{black}{
 We propose an online multi-agent QL algorithm integrating the above action selection policies. 
 The proposed online QL algorithm utilizes a reward function that considers a handover cost while maximizing the network sum rate.} 
 The signaling overhead and computational complexity of this proposed QL algorithm are analyzed in detail for both the centralized and distributed versions, showing significantly lower overhead for the distributed implementation.


\item 
We perform extensive numerical evaluations in both static and dynamic network settings to assess the convergence rate, performance, and adaptability of the proposed QL algorithms. Results show significant improvements in both throughput and handover rate over the 3GPP max-SINR association method, reaching close to the near-optimal WCS sum rate at a much lower handover frequency.
Results also show that the proposed online multi-agent QL algorithms are robust and can adapt well to the dynamics of mmWave channel variation and user mobility under various UE speed profiles including walking, running, biking and suburban driving, validating their use in 
real-time wireless network.
\end{itemize}

\textcolor{black}{The rest of this paper is structured as follows. Section II introduces the system model including channel model and UE mobility model, followed by the formulation of an optimization problem for load balancing user association.
We present the multi-agent model for Q-learning in Section III. Next, we propose the centralized and distributed multi-agent policies for load balancing in Section IV and Section V. 
Section VI details our online multi-agent QL algorithm for handover utilizing the proposed policies, followed by an analysis of signaling overhead and complexity in Section VII.
We evaluate performance of the proposed multi-agent QL algorithm via simulation in Section VIII and provide our conclusions in Section IX.
}


\section{System Models}
\textcolor{black}{In this section, we introduce the system models, including the network, channel, and signal models, and define the association variables. 
Furthermore, we discuss a UE mobility model and frame measurement structure which will enable multi-agent Q-learning to adapt to mobile UEs.}
While we consider a multi-tier system in this paper, the subsequent proposed learning algorithms can be applied in either a single-tier or multi-tier network.

\subsection{Network and Channel Models}
\label{Sys_Ch_models}

\begin{figure}[t]
  \centering
\includegraphics[width=85mm]{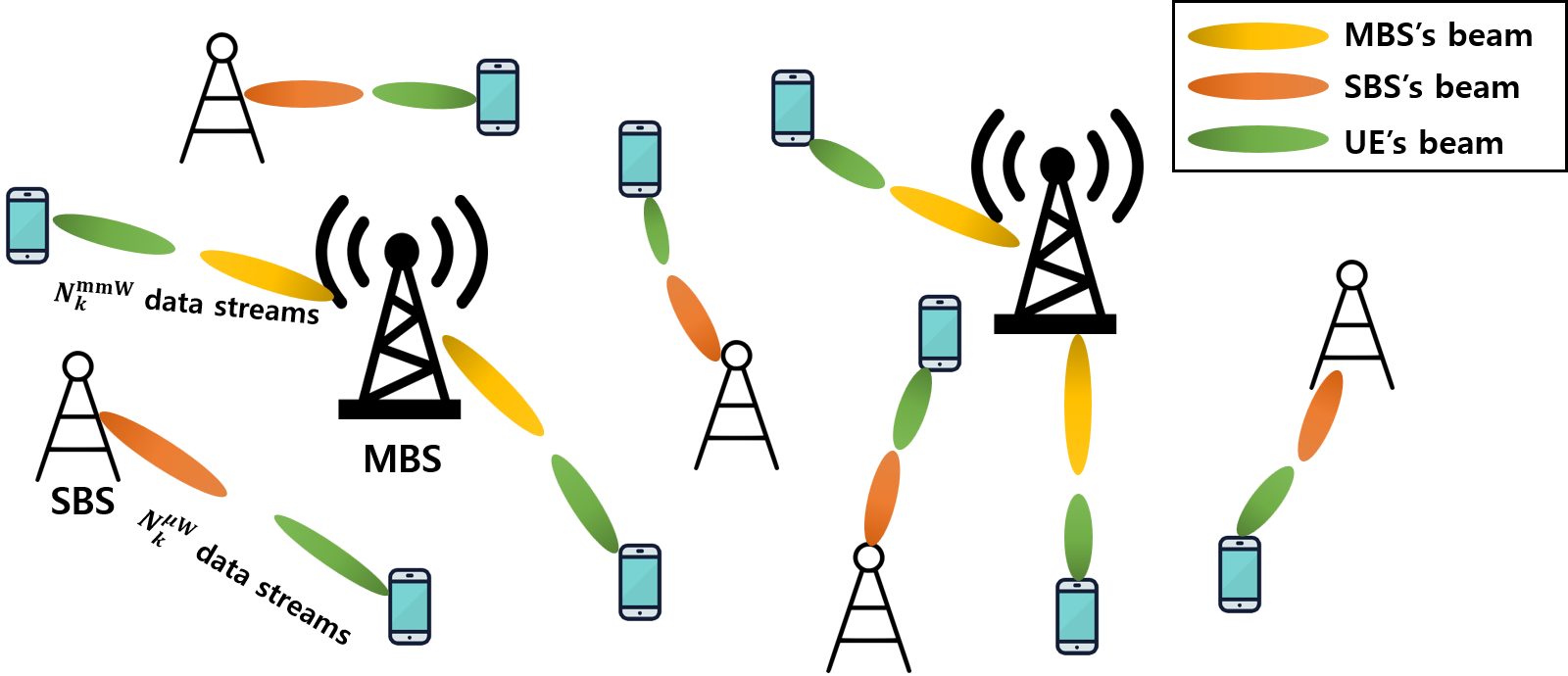}
\caption{\textcolor{black}{Illustration of a two-tier cellular HetNet.
Each user is associated with one of $J_M$ MBSs and $J_s$ SBSs while requesting $N_k^\text{mmW}$ data streams from a MBS or $N_k^{\mu \text{W}}$ data streams from a SBS.}}
\label{fig:system}
\end{figure}

\textcolor{black}{
We consider the downlink of a two-tier cellular network including $J_M$ MBSs in tier-1, $J_S$ SBSs in tier-2, and $K$ UEs as shown in Fig. \ref{fig:system}, where the MBSs and SBSs operate at sub-6GHz and mmWave, respectively.} 
Let $\mathcal{J}= \{1, ..., j, ..., J\}$ denote the set of all BSs with $J=J_M+J_S$, and $\mathcal{K}=\{1, ..., k, ..., K\}$ represents the set of UEs. 

Each BS $j$ is equipped with an antenna array of size $M_j$ and has a specific \textit{quota} $m_j$ representing the maximum number of downlink data streams it can transmit. Thus, the \textit{quota vector} of BSs can be defined as $\mathbf{m}=[m_1,...,m_J]$. We assume $1\leq m_j\leq M_j$, where the upper bound is due to the fact that the number of data streams transmitted by each BS cannot exceed the number of its antennas. 
\textcolor{black}{
Each UE is equipped with two antenna arrays: 1) an array with $N_k^{\mu \text{W}}$ elements for sub-6 GHz band, and 2) an array with $N_k^\text{mmW}$ elements for mmWave band.
As such, each UE $k$ can request up to $n_k$ data streams from its serving BS where $n_k\in \{N_k^\text{mmW}, N_k^{\mu \text{W}}\}$ depending on the BS type as shown in Fig. \ref{fig:system}.}

The Gaussian MIMO channel $\mathbf{H}^{\mu\text{W}}_{k,j}\in \mathbb{C}^{N_k^{\mu \text{W}}\times M_j}$ with zero-mean and unit variance i.i.d. complex Gaussian random variables is used for sub-6 GHz transmissions from MBSs. In the mmWave band, however, the transmissions are highly directional due to beamforming at both transmitter and receiver sides and we employ the clustered mmWave MIMO channel $\mathbf{H}^{\text{mmW}}_{k,j}\in \mathbb{C}^{N_k^{\text{mmW}}\times M_j}$ \cite{3GPP901}, \cite{Nokia}.
We use the probability of LoS and NLoS and the path loss model for LoS and NLoS transmissions as given in \cite{3GPP901}. 

\subsection{User Association and Data Rate}

\textcolor{black}{
The association vector $\boldsymbol{\upeta}^{(t)}$ specifies the connections between UEs and BSs at time step $t$ and is defined as
\begin{equation}
\label{upeta}
\boldsymbol{\upeta}^{(t)}\triangleq[{\eta}_1^{(t)}, ..., {\eta}_{K}^{(t)}]^T,
\end{equation}
where ${\eta}_k^{(t)}\in\mathcal{J}$ represents the index of BS associated with user $k$.
}

\textcolor{black}{
Each BS $j$ can transmit a maximum number of $m_j$ data streams to the associated UEs. We formulate this quota by defining a \textit{load balancing} constraint as
\begin{equation}
\sum_{k\in\mathcal{K}} n_k \mathds{1}_{k,j}^{(t)} \leq m_j,
\label{LBC}
\end{equation}
where $\mathds{1}_{k,j}^{(t)}=1$ if $\eta_{k}^{(t)}=j$, and $\mathds{1}_{k,j}^{(t)}=0$ if $\eta_{k}^{(t)}\neq j$. 
}

\textcolor{black}{
Highly directional beamforming transmissions and the fast-varying nature of mmwave channels make interference highly dependent on user association \cite{TWC}. 
We consider this dependency and formulate the data rate (reward) of UE $k$ from BS $j$ at time step $t$ as \cite{TWC_alireza}
\begin{align}\label{R_kvj}
  &R_{k,j}\bigl(\boldsymbol{\upeta}^{(t)}\bigr) 
 \\\notag
 &~= \log_2\left[\text{det}\left(\mathbf{I}_{n_k}+\left(\mathbf{V}_{k,j}\bigl(\boldsymbol{\upeta}^{(t)}\bigr)\right)^{-1}\mathbf{W}_k^*\mathbf{H}_{k,j}\mathbf{F}_{k,j}\mathbf{F}_{k,j}^*\mathbf{H}_{k,j}^*\mathbf{W}_{k} \right)\right],
\end{align}
where $\mathbf{V}_{k,j}\bigl(\boldsymbol{\upeta}^{(t)}\bigr)$ is the interference-plus-noise covariance matrix given as
\begin{align}\label{Eq:cov_mat}
\mathbf{V}_{k,j}\bigl(\boldsymbol{\upeta}^{(t)}\bigr)=&\mathbf{W}_{k}^*\mathbf{H}_{k,j}\biggl(\sum_{\substack{l\in \mathcal{K}_j^{(t)},l\neq k}}\mathbf{F}_{l,j}\mathbf{F}_{l,j}^*\biggr)\mathbf{H}_{k,j}^*\mathbf{W}_{k}\\\notag
+&\mathbf{W}_{k}^*\biggl(\sum_{\substack{i\in \mathcal{J}, i\neq j}} \sum_{l\in \mathcal{K}_i^{(t)}} \mathbf{H}_{k,i}\mathbf{F}_{l,i}\mathbf{F}_{l,i}^*\mathbf{H}_{k,i}^*\biggr)\mathbf{W}_{k}+N_0\mathbf{W}_{k}^*\mathbf{W}_{k}.
\end{align}
Here $\mathbf{F}_{k,j}\in\mathbb{C}^{M_j\times n_k}$ is the linear precoder (transmit beamforming matrix) at BS $j$ intended for UE $k$, and $\mathbf{W}_{k}\in\mathbb{C}^{N_k \times n_k}$ is the linear combiner (receive beamforming matrix) of UE $k$. 
Set $\mathcal{K}_{j}^{(t)}$ is the activation set of BS $j$ at time $t$, which includes all UEs associated with the BS at that time. 
Eq. \eqref{Eq:cov_mat} shows the dependency of interference on user association. 
We note that there is no inter-tier interference as the two tiers are operating at two separate frequency bands.}

\textcolor{black}{
The total learning reward at each time step $t$ is computed as the overall \textit{network sum rate} given by
\begin{align}\label{r_t}
r\bigl(\boldsymbol{\upeta}^{(t)}\bigr)=\sum_{j\in \mathcal{J}}\sum_{k \in \mathcal{K}} R_{k,j}\bigl(\boldsymbol{\upeta}^{(t)}\bigr),
\end{align}
which will also be used as a measure of network performance. 
}

\subsection{User Mobility Model}
\label{Mob_Meas_Model}
In a mobile network, UEs are moving and keeping track of the best associations is challenging. In a mmWave-enabled network, this problem is even more challenging since mmWave channel conditions can change abruptly in short periods of time. Next, we consider a mobility model and introduce a measurement model which will be used in Sec. \ref{UA_HO_Sec} to study handover in such a highly-dynamic network.

For UE movements, we utilize the \textit{modified random waypoint} (MRWP) model, which is defined by an infinite sequence of quadruples as $\{(\mathbf{X}_{k,n-1},\mathbf{X}_{k,n},V_{k,n},Z_{k,n})\}_{k\in\mathcal{K},~n\in\mathbb{N}}$, where $n$ denotes the \textit{moving step} $n$ during which UE $k$ travels from starting waypoint $\mathbf{X}_{k,n-1}$ to target waypoint $\mathbf{X}_{k,n}$ \cite{MRWP_Mobility}. In this mobility model, each UE $k$ uniformly selects a random velocity $V_{k,n}\in (0, V_\text{max}]$, and a random pause time $Z_{k,n}$ at the target waypoint. Then, given a source waypoint $\mathbf{X}_{k,n-1}$, the UE generates a homogeneous Poisson point process (PPP) $\Phi(n)$ with intensity $\lambda$, and selects the nearest point in $\Phi(n)$ as the target waypoint, i.e., $\mathbf{X}_{k,n}=\argminA_{\mathbf{x}\in\Phi(n)}||\mathbf{x}-\mathbf{X}_{k,n-1}||$. Thus, the \textit{transition length} of UE $k$ during moving step $n$ can be calculated as $L_{k,n}=||\mathbf{X}_{k,n}-\mathbf{X}_{k,n-1}||$,
and its \textit{transition time} is $T_{k,n}=L_{k,n}/V_{k,n}$.

\begin{figure}[t]
\vspace*{-1em}
\centering
\includegraphics[width=90mm]{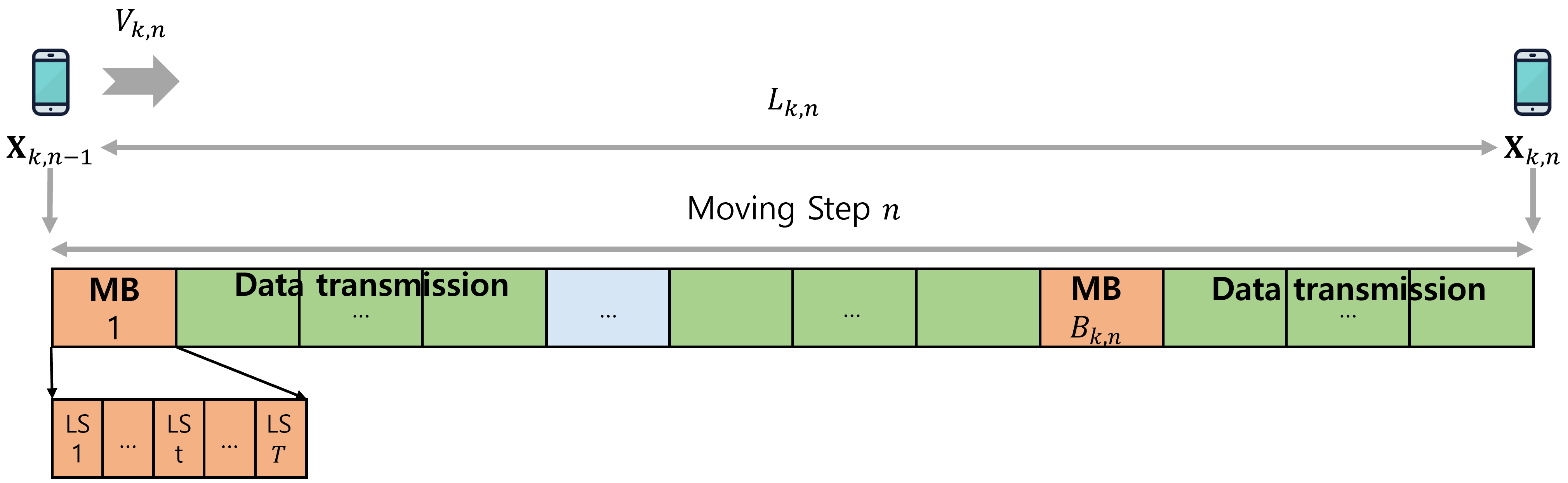}
\vspace*{-1em}
\caption{Structure of moving step $n$ during which UE $k$ travels from source waypoint $\mathbf{X}_{k,n-1}$ to target waypoint $\mathbf{X}_{k,n}$ with velocity $V_{k,n}$. The number of MBs for each moving step is obtained according to (\ref{N_MB}), which depends on UE velocity $V_{k,n}$, distance between source and target waypoints $L_{k,n}$, and time duration of each MB $t^\text{MB}$.}
\label{moving_step}
\end{figure}

We assume each moving step is composed of multiple \textit{measurement blocks} (MBs), 
interspersed between data transmission blocks as shown in Fig. 1.
Inside each MB, each UE performs a set of measurements from its serving and neighbor BSs. 
The measurement quantities, as specified in the 3GPP standards, can be the reference signal received power (RSRP), reference signal received quality (RSRQ), or SINR \cite{3GPP331}. 
These measurements are then reported to the network every $t^\text{MB}$ seconds\footnote{This parameter is known as information element \textit{ReportInterval} in 3GPP standardization, which indicates the interval between periodical reports. The range of value for the ReportInterval in 5G NR is 120 ms - 30 min \cite{3GPP331}.} for the purpose of handover. 
Thus, the number of MBs for UE $k$ during moving step $n$ is determined by rounding up the ratio between the transition time of the UE and the report interval as follows:
\begin{equation}
\label{N_MB}
B_{k,n}=\left\lceil\frac{T_{k,n}}{t^\text{MB}}\right\rceil.
\end{equation}
Thus, $B_{k,n}$ is random and changes from one moving step to another. At the end of each MB, the network can make handover decisions based on the new measurements, which may lead to changes in associations.

\subsection{Optimization Problem}

The goal is to find the optimal association vector $\boldsymbol{\upeta}^{(t)}$ at each learning step, which maximizes a network utility function. In this paper, we consider network throughput (\ref{r_t}) as the utility function and define the user association problem at learning step $t$ as
\begin{subequations}\label{opt_prob2}
\begin{align}
\maxi_{\boldsymbol{\upeta}^{(t)}}~~&r(\boldsymbol{\upeta}^{(t)})\\
\mathrm{subject~to}~~&\sum_{j\in\mathcal{J}} \mathds{1}_{k,j}^{(t)} \leq 1, ~~\forall k\in \mathcal{K}\\
&\sum_{k\in\mathcal{K}} n_k \mathds{1}_{k,j}^{(t)} \leq m_j, ~~\forall j\in \mathcal{J}
\end{align}
\end{subequations}
The set of constraints in (\ref{opt_prob2}b) represents the unique association constraints, meaning that each UE can only connect to one BS at a time. The constraints in (\ref{opt_prob2}c) allows our association scheme to limit each BS's load separately. The load balancing quota  ($m_j$) is set by the available resources at each BS, and this set of constraints guarantees that each BS can serve all its associated UEs up to the specified load.



Under load balancing constraints, the associations of all UEs are inter-dependent. 
This problem can be solved by a centralized algorithm such as WCS \cite{TWC}, which make association decisions by considering the dependency among UEs while satisfying load balancing constraints.
However, a centralized optimization algorithm requires significant computational complexity and high signaling overhead such as delivering the measured SINR values and global channel state information (CSI) to a central server. 
Moreover, performing optimization in a highly dynamic network may be impractical, since the optimization needs to be re-solved at each measurement block $b$ to effectively adapt to the network dynamics. 
In this paper, we propose a multi-agent reinforcement learning approach for user association, which can reduce signaling overhead by using only local information at each user and adapting in real-time to changes in the environment.

\section{Multi-agent QL Model For User Association}


User association can be cast as a multi-agent QL (QL) problem in which UEs (or UEs) are agents, selecting a BS corresponds to taking an action, the wireless networks represents the environment, and maximizing the network utility is the long-term goal. In a multi-agent QL system, ideally the agents take actions independently using only local information. 
The load balancing constraints in (\ref{opt_prob2}c), however, introduce an important dependency among the actions taken by different agents.

There have been MARL (multi-agent reinforcement learning) architectures which deal with the dependency among different agents' actions. These include introducing communications among agents such as CommNet \cite{sukhbaatar2016learning} and IC3Net \cite{singh2018learning}, or using Actor-Critic-type algorithms with centralized training and decentralized execution such as MADDPG \cite{lowe2017multi} and MAAC \cite{iqbal2019actor}. The former approach of introducing communications among agents would require device-to-device communications among UEs, which we do not assume here. The latter approach of centralized training and decentralized execution allows agents to observe some implicit dependency, but does not guarantee that a constraint such as load balancing can be met at every execution step.

In this paper, we propose a multi-agent QL system which guarantees load balancing at every execution step. The proposed QL system employs a multi-agent action selection policy to ensure load balancing, which is described in the subsequent two sections. 
\textcolor{black}{
Next, we define the parameters of the QL model at each agent, and introduce the Q-value update rule at each agent and the multi-agent actions selection metric based on the upper confidence bound (UCB).
}


\subsection{Cellular Network QL Components}
The components of our proposed QL-based user association model are defined as follows. 
Each UE represents an \textit{agent} which interacts with the cellular network to achieve a goal. The considered QL model therefore is a multi-agent system with $K$ agents. 
Each UE $k$ perceives its environment through a set $S_k$ consisting of states of the form $s_k=\left(\eta_k,\mathbold{\rho}_k\right)\in\mathcal{S}_k$, where $\eta_k\in\mathcal{J}$ is the association variable specifying the BS serving (connecting with) UE $k$, and $\mathbold{\rho}_k$ is a vector of size $J$ denoting the quantized SINR values from UE $k$ to the $J$ considered BSs, given current connection state with BS $\eta_k$. 
The $\eta_k^\text{th}$ element of $\mathbold{\rho}_k$ represents the quantized SINR level of the BS associated with UE $k$, and all other elements denote the SINR levels of other BSs not associated with UE $k$ in this state. 
In practice, SINR values are measured for each data stream of UE $k$, and vector $\mathbold{\rho}_k$ is obtained by averaging the SINR values over all data streams at this UE.

We use two sets of SINR quantization levels: $S$ values ($S > 2$) for the associated BS where the quantization is uniform between an $\text{SINR}^\text{min}$ and an $\text{SINR}^\text{max}$ value, and binary quantization (2 values) for other BSs as these are less sensitive for performance. 
As a result, the size of state space for each UE is $|\mathcal{S}_k|= 2^{J-1}JS$.

At each learning step $t$, UE $k$ realizes a specific state $s_k^{(t)} = \left(\eta_k^{(t)}, \rho_k^{(t)}\right) \in \mathcal{S}_k$. The association vector $\boldsymbol{\upeta}^{(t)} = \left[\eta_1^{(t)}, ..., \eta_K^{(t)}\right]$ contains the current association states of all UEs. 
At each step $t$, UE $k$ takes a specific action $a_k^{(t)}\in\mathcal{J}$ and transitions from state $s_k^{(t)}$ to state $s_k^{(t+1)}=\left(\eta_k^{(t+1)},\rho_k^{(t+1)}\right)$, where  $\eta_k^{(t+1)}=a_k^{(t)}$.
After taking an action $a_k^{(t)}$, each UE $k$ receives an instantaneous reward $R_{k,a_k^{(t)}}$ as in (\ref{R_kvj}), which will be used to determine its next SINR state and update its Q-value.

\subsection{Measurement Model and Learning Steps}
\textcolor{black}{In this subsection, we define the learning process and association process by utilizing a frame structure that contains multiple measurement blocks as in Fig. \ref{moving_step}.}
Recall from the mobility model that each UE's moving step contains multiple measurement blocks as defined in \eqref{N_MB}. 
\textcolor{black}{
Each measurement block (MB) $b$ contains a fixed number of learning steps, denoted as $T$. 
We assume that channel state information (CSI) is static within an MB, but dynamically changes in the next MB.
At the beginning of each MB, each UE (agent) $k$ takes a measurement of its SINR vector $\rho^{(b)}_k$. 
During learning step $t$, each agent obtain its current state $s_k^{(t)}$, chooses an action, and update its chosen Q-value (as discussed in the next subsections).
Even though the CSI is static during each MB consisting of $T$ learning steps, the current chosen Q-value of each UE will change in each learning step as shown later in the updating rule \eqref{QL_UR}, thereby facilitating the learning process.
}

\textcolor{black}{
The actions from all agents at each learning step $t$ produce a new learning association vector $\boldsymbol{\upeta}^{(t)}$, which will be updated through the multi-agent policy described in Sections IV and V to ensure load balancing at every BS.
This learning association vector, however, is not used for performing association and handover. Instead, we keep another association vector $\boldsymbol{\upbeta}^{(b)}$, the \emph{best-to-date association}, 
which is updated at the end of each measurement block $b$, after every $T$ learning steps, and which will be used for performing connections (association and handover). 
We note that while it is also possible to update the best-to-date association and perform connections at the end of each learning step, this may result in less stable associations and more frequent handovers.
}

\textcolor{black}{
We note also that the assumption of fixed CSI per measurement block is not essential for our QL algorithms to work. In a fast fading environment, it is possible to have CSI varied during an MB, in which case each UE will perform multiple measurements during each MB. Specifically, each UE can perform one measurement of its $\rho_k^{(t)}$ per learning step, for a total of $T$ measurements in each MB. The role of $T$ in both cases of slow and fast fading is to allow the update of best-to-date association vector after a number of learning steps, rather than at every learning step to reduce handover frequency.
}

Specifically, each measurement block $b$ has an 
association vector $\boldsymbol{\upbeta}^{(b)}$, chosen from 
all learning association vectors $\boldsymbol{\upeta}^{(t)}$ within that measurement block as
\begin{equation}
\boldsymbol{\upbeta}^{(b)}= \argmaxA_{t=\{1,...,T\}}  r(\boldsymbol{\upeta}^{(t)}),
\label{best_assoc}
\end{equation}
where $r(.)$ is given in (\ref{r_t}) and the initial learning association vector at each MB $b$ is the best-to-date association vector from the previous MB $b-1$, that is $\boldsymbol{\upeta}^{(1)}=\boldsymbol{\upbeta}^{(b-1)}$. In this way, $\boldsymbol{\upbeta}^{(b)}$ is the best-to-date  association vector taking into account all past history.



By using these two different association vectors, we distinguish between the learning process and the association process. This is done by performing two separate updating procedures: 1) behavioral update (or learning update of $\eta_k^{(t)}$), and 2) target update (or operation/association update of $\beta_k^{(b)}$). 
The behavioral procedure updates $\eta_k^{(t)}$ at each learning step (as discussed later in \eqref{QL_UR}), adapting to the immediate changes in the environment, whereas the target procedure updates $\beta_k^{(b)}$ only when the new association vector results in a higher network sum as in \eqref{best_assoc}, thus aiming at increasing the long-term expected reward. In a sense, our update of the best-to-date association vectors is similar to the target Q-network update in \cite{mnih2015human},  
except that $\beta_k^{(b)}$ is updated as best-to-date instead of periodically, and is used for performing the actual association connection instead of learning.

\subsection{{Q-value Update at Each UE}} 
\textcolor{black}{
To focus our attention on the multi-agent policies design (Sections IV and V), we choose standard tabular-based Q-value update in this paper.
The proposed multi-agent action selection policies, however, are also applicable to deep Q-networks (DQN), which can handle a large state-action space. In other words, our Q-learning framework can be applied with any training and updating method, including either standard-QL \cite{sutton2018reinforcement} or  deep Q-learning \cite{mnih2015human,van2016deep}. 
}

In standard QL, each UE $k$ locally maintains and updates a \textit{Q value table} $\mathbf{Q}_k$ containing the Q-values for all its state-action pairs. The table has the size of $|\mathcal{S}_k| \times J$ (total number of states $\times$ number of actions). 
Each entry of this Q-table is called a {Q-value} $Q_k(s_k,a_k)$ which is the current estimate of the long-term expected reward for taking action $a_k$ in state $s_k$.
\textcolor{black}{Since the size of the state-action space increases exponentially with the number of BSs, for large networks, each Q-table can be replaced by a deep Q-network (DQN). In DQN, a DNN (deep neural network) is used at each UE to compute the Q-values for all possible actions given an input state.
Since the input size only grows linearly with the number of BSs, DQN can handle a large state-action space. We will not discuss DQN implementation in this paper, but note that our proposed multi-agent policies (Sections IV and V) and online algorithm (Section VI) can be applied to DQNs as well.}


At each learning step $t$, when user $k$ in current state $s_k^{(t)}$ takes an action $a_k^{(t)}$, the value for this state-action pair is updated in this UE's Q-table as follows \cite{ML_Intro_Book}:
\begin{align}
Q_k\left(s_k^{(t+1)},a_k^{(t+1)}\right)=&~(1-\alpha)Q_k\left(s_k^{(t)},a_k^{(t)}\right)\nonumber\\+&\alpha\left[\tilde{R}_{k,a_k}+\gamma \max_{b_k\in \mathcal{J}} Q_k\left(s_k^{(t+1)},b_k\right)\right],
\label{QL_UR}
\end{align}
where $\alpha\in[0,1)$ is the learning rate, 
$\gamma\in[0,1)$ is the discount factor for future rewards, $b_k$ represents all possible actions in the next state, among which the best is chosen, and $\tilde{R}_{k,a_k}$ is the immediate reward for agent $k$ by taking action $a_k$ in the current state. The choice of the reward function is important and will be discussed in Section \ref{UA_HO_Sec}.

\subsection{Multi-agent Actions Selection Metric}
\label{UCB_Sec}
In order to determine which entry of the Q-table of each UE to be updated as in \eqref{QL_UR}, we need a \textit{policy} which defines how an UE $k$ chooses the action $a_k^{(t)}$ given its current state. The policy needs a metric to select the actions.

Different metrics have been used in the literature for deciding the action at an agent. Here we choose the UCB (upper confidence bound), instead of the direct Q value, as the metric for action selection. The UCB metric allows a balance between \textit{exploiting} the current best action and \textit{exploring} other actions \cite{ML_Intro_Book}.  
For each UE $k$ at current state $s_k^{(t)}$, the UCB value (or a U-value) for a state-action pair is defined as follows 
\begin{equation}
\label{UCB_rule}
{U}_k\left(s_k^{(t)},a\right) = Q_k\left(s_k^{(t)},a\right)+c\sqrt{\frac{\ln t}{N_k^{(t-1)}\left(s_k^{(t)},a\right)}},
\end{equation}
where $c>0$ controls the degree of exploration, and $N_k^{(t-1)}\left(s_k^{(t)},a\right)$ is the number of times action $a$ has been selected at state $s_k^{(t)}$ up to and including learning step $t-1$. Each time an action $a$ is taken by UE $k$, this number is updated as follows
\begin{equation}\label{NBS}
N_k^{(t)}\left(s_k^{(t)},a\right) = N_k^{(t-1)}\left(s_k^{(t)},a\right) +1.
\end{equation}
In the next sections, we will discuss how the U-values in \eqref{UCB_rule} are updated and used in the multi-agent actions selection policy to ensure load balancing at every step.

\section{Centralized Multi-agent Actions Selection Policy for Load Balancing}
\label{LBA_Sec}

\textcolor{black}{
In the centralized approach, a central entity determines the actions for all agents to ensure load balancing at every BS. 
We propose the use of a central entity called the {\it CLB} to manage actions selection. 
The CLB runs an algorithm to maximize the total UCB values for all UEs while satisfying the load balancing constraints in (\ref{LBC}). 
This CLB produces a load-balanced association vector $\boldsymbol{\upeta}^{(t)}$ at each learning step and sends the corresponding association result to each UE.
}

\label{WCS_CLB_Sec}

\subsection{CLB Record Keeping and Updates}
In order to perform centralized load balancing, the CLB needs to collect the Q-values from all UEs for all state-action pairs. This collection needs to occur \emph{only once} at the initialization of the learning process, during which each UE sends its Q-table to the CLB. After this initialization, at each learning step, each UE reports a single Q-value to the CLB for the specific state-action pair updated at the UE in that step as in (\ref{QL_UR}). The CLB then updates and maintains its record of the Q-tables from all UEs. Note also that each UE also maintains and updates its own Q-table. 
In other words, the CLB keeps a record of $K$ Q-tables of all $K$ UEs, and updates one entry, $Q_k\left(s_k^{(t)},a_k^{(t)}\right)$, in each table at each learning step, when that entry is also updated in the Q-table at the corresponding UE. We emphasize that the Q-values are not shared among agents, but are only shared with the CLB, facilitating a multi-agent QL without information exchange among agents.


Similarly, the CLB maintains a set of N-tables of all $K$ UEs, which is initialized to 0 at the beginning of the learning process. Note that the N-tables are only kept at the CLB but not at the UEs, and there is no communication between UEs and the CLB on the N-values. At each learning step, the CLB updates one entry of each N-table according to the state-action pair of the corresponding UE as in (\ref{NBS}).

At each learning step, the CLB extracts a row from each of its Q-tables and N-tables record to compute a network U-table corresponding to the UEs' current state $\mathbf{s}^{(t)} = \left[s_1^{(t)}, ... s_k^{(t)}, ..., s_K^{(t)}\right]$. Specifically, the CLB forms the network U-table of size $K\times J$ consisting of one row for each UE as follows:
\begin{equation}
{\mathbf{U}}(\mathbf{s}^{(t)})=\left[{\mathbf{u}_1^T}\left(s_1^{(t)}\right); ... {\mathbf{u}}_k^T\left(s_k^{(t)}\right); ...{\mathbf{u}}_K^{T}\left(s_K^{(t)}\right)\right],
\label{U-table}
\end{equation}
where the U-values are computed using the Q-values and N-values as in (\ref{UCB_rule}).
The CLB then uses this network U-table to perform load balancing assignment, using \emph{the CLB load balancing algorithm} proposed below, to obtain a new learning association vector. Then, the CLB informs all the UEs about their load-balanced association, which determines their action to be taken.

When a UE carries out this action, it receives a reward, updates its Q-value, and reports this value to the CLB. Then, the CLB  updates the corresponding entry in the Q-table and N-table of that UE.

\begin{figure}
    \centering
    \includegraphics[width=90mm]{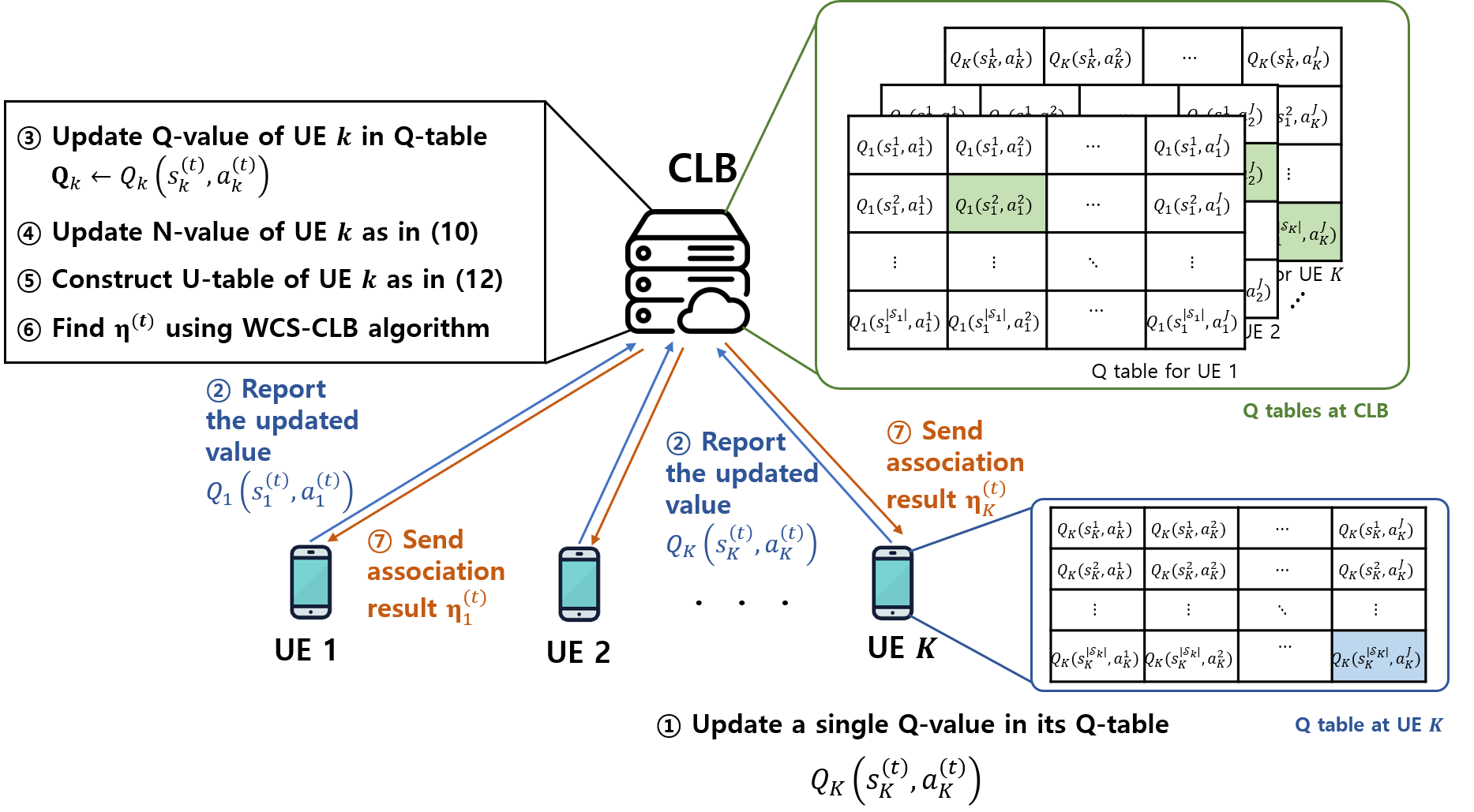}
    \caption{\textcolor{black}{Procedure for Q-value update at the CLB in the proposed centralized action selection policy}}
    \label{fig:new_cent}
\end{figure}

\subsection{CLB Load Balancing Action Selection Policy}

\textcolor{black}{
As discussed above, in each learning step $t$, the CLB receives an updated Q-value from each UE. The CLB uses each reported value to update its Q-table for the corresponding UE, and also calculate the update in its N-table for that UE. The UEs need not send any N-value. Next, the CLB produces the U-table as in \eqref{U-table}, which contains the U-values of the current learning step for all possible connections. Then, the CLB aims to find the association vector which achieves the maximum sum U-value for this learning step.}
This problem is an integer programming, whose optimal solution would require an exhaustive search which is exponential in complexity. We use the idea of WCS in \cite{TWC} which is a polynomial-time method to achieve a high sum U-value at the CLB while satisfying load balancing. 
The principal idea of WCS is to start with an association vector that is load-balanced, then swap the worst connection with another connection to improve the U-value sum while maintaining load balancing. The worst connection under an association vector $\boldsymbol{\upeta}^{(i)}$ is the one with lowest U-value in ${\mathbf{U}}^{(t)}$. For each swapping step $i$, we find a new association vector $\boldsymbol{\upeta}^{(i+1)}$ to maximize sum U-value as follows.
\begin{equation}\label{eq:WCS_CLB}
\boldsymbol{\upeta}^{(i+1)}=\argmaxA_{\{\boldsymbol{\upeta}^{(i)}_{n} | n\in{\mathcal{K}},~n\neq k^{(i)}\},~\boldsymbol{\upeta}^{(i)}} ~\sum_{k\in\mathcal{K}} {\mathbf{U}}_{k,\eta_{k}^{(i)}}^{(t)}.
\end{equation}
where $\boldsymbol{\upeta}^{(i)}_{n}$ is the new association vector by swapping the worst connection UE $k^{(i)}$ with the connection of UE $n$.
In addition, we apply a \textit{switching step} to explore other possibilities for improving the objective function. If $\boldsymbol{\upeta}^{(i+1)}=\boldsymbol{\upeta}^{(i)}$, we compute a UE switching index $l^{(m)}=\text{mod}(m-1,K)+1$
and switch the $k^{(i+1)^\text{th}}$ and $l^{(m)^\text{th}}$ elements of the
current association vector to obtain the switched association vector $\boldsymbol{\upeta}^{(i+1)}_\text{sw}$. 
This new association vector is used as the initial association vector for the next swapping iteration. The process repeats until no improvement in the U-value sum is found. 
The proposed algorithm guarantees convergence because the objective in \eqref{eq:WCS_CLB} is always increasing. The detailed WCS-CLB algorithm is summarized in Alg. \ref{WCS_CLB_Alg}.
\textcolor{black}{
Fig. 3 illustrates the complete procedure for Q-value updates and association vector selection at the CLB.
}

\textcolor{black}{
\textbf{\textit{Remark 1:}} 
The CLB-WCS algorithm starts with an initial association vector which is chosen to satisfy load balancing at all BSs. Then during the process, each swapping or switching step does not change the load at any BS, but simply swaps or switches the UEs associating with two different BSs. Thus the resulting association vector at every step of the CLB-WCS algorithm always satisfies load balancing at all BSs. That is, by starting with a load-balanced initial association, CLB-WCS guarantees an output association vector satisfying all load balancing constraints.
}

\textcolor{black}{
\textbf{\textit{Remark 2:}} During the execution of the WCS-CLB algorithm at each learning step, the network U-table ${\mathbf{U}}\left(\mathbf{s}^{(t)}\right)$ remains fixed at the CLB and no communications between UEs and the CLB occur. Communications between UEs and the CLB occur only  before and after each execution of the WCS-CLB. Before each run of the WCS-CLB, each UE sends a single Q-value to the CLB for updating and computing the U-table in \eqref{U-table}. After each WCS-CLB execution, the CLB delivers the association results $\{\eta^{(t)}_k\}$ by sending a single associated BS index to each UE. As such, the communication requirement between UEs and the CLB is minimal.
}

\begin{algorithm}[t]\footnotesize
\SetAlgoLined
\KwInput{Q values $Q_k(s_k^{(t)},a_k^{(t)}), \forall k\in\mathcal{K}$ reported from all UEs, and quota vector $\mathbf{m}$\\
\textbf{Initialization}:}
~~~- Initialize $i=1$ and $m=1$\;
~~~- The CLB forms the network U-table $\tilde{\mathbf{Q}}^{(t)}\left(\mathbf{s}^{(t)}\right)=\left[\tilde{\mathbf{q}}_1^{(t)}\left(s_1^{(t)}\right),...,\tilde{\mathbf{q}}_K^{(t)}\left(s_K^{(t)}\right)\right]$\;
~~~- Generate an arbitrary feasible association vector $\boldsymbol{\upeta}^{(1)}$ according to $\mathbf{m}$\;
~~~- Find the initial worst connection $\left(k^{(1)},j^{(1)}\right)=\argminA_{k,j} \tilde{\mathbf{Q}}_{k,\eta_{k}^{(1)}}^{(t)}$\;
\While{stopping criterion is not met}{
\textit{Swap} the element $k^{(i)}$ with all other elements of $\boldsymbol{\upeta}^{(i)}$ to obtain new vectors $\boldsymbol{\upeta}^{(i)}_{n}$\; 
Find the best association vector $\boldsymbol{\upeta}^{(i+1)}=\mathrm{arg}~\max_{\{\boldsymbol{\upeta}^{(i)}_{n} | n\in{\mathcal{K}},~n\neq k^{(i)}\},~\boldsymbol{\upeta}^{(i)}} ~\sum_{k\in\mathcal{K}} \tilde{\mathbf{Q}}_{k,\eta_{k}^{(i)}}^{(t)}$\;
Find the worst connection $\left(k^{(i+1)},j^{(i+1)}\right)=\argminA_{k,j} \tilde{\mathbf{Q}}_{k,\eta_{k}^{(i+1)}}^{(t)}$\;
\If{$\boldsymbol{\upeta}^{(i+1)}=\boldsymbol{\upeta}^{(i)}$}{
$l^{(m)} \leftarrow \text{mod}(m-1,K)+1$\;
\textit{Switch} the elements $k^{(i+1)}$ and $l^{(m)}$ of $\boldsymbol{\upeta}^{(i+1)}$ to obtain $\boldsymbol{\upeta}^{(i+1)}_\text{sw}$\;
$\boldsymbol{\upeta}^{(i+1)} \leftarrow \boldsymbol{\upeta}^{(i+1)}_\text{sw}$\;
$m \leftarrow m+1$\;
}
\textbf{Stopping criterion}: Break if the best $\boldsymbol{\upeta}$ does not change after $K$ consecutive iterations\;
$i \leftarrow i +1$\;
}
Set the learning association vector $\boldsymbol{\upeta}^{(t)}=\boldsymbol{\upeta}^{(i)}$\;
CLB informs each UE $k$ about its $\eta_k^{(t)}$\;
\KwOutput{Learning association vector $\boldsymbol{\upeta}^{(t)}$}
\caption{(WCS-CLB) Centralized Action Selection Policy}
\label{WCS_CLB_Alg}
\end{algorithm}

 
\begin{algorithm}[t]\footnotesize
\SetAlgoLined
\KwInput{Q values $Q_k(s_k^{(t)},a_k^{(t)}), \forall k\in\mathcal{K}$ reported from all UEs, and quota vector $\mathbf{m}$\\
\textbf{Initialization}:}
~~~- Set $n=1$ and $l_k=1,~\forall k$\;
~~~- Each UE $k$ builds its preference lists ($\mathcal{P}_k^\text{UE}$) using its U-vector $\tilde{\mathbf{q}}_k^{(t)}$\;
~~~- Each BS $j$ builds its preference lists ($\mathcal{P}_j^\text{BS}$) using its U-values $\tilde{Q}_k\left(s_k^{(t)},j\right)$\;
~~~- Form the initial rejection set $\mathcal{R}=\{1, 2, ..., K\}$\;
~~~- Initialize the waiting list of each BS $\mathcal{W}_j^0=\varnothing,~\forall j$\;
~~~- Perform a DA matching game \cite{gale1962college} as follows:\\
\While{$\mathcal{R}\neq\varnothing$}{
Each UE $k\in\mathcal{R}$ applies to its $l_k^\text{th}$ preferred BS in $\mathcal{P}_k^\text{UE}$\;
Each BS $j$ forms its current waiting list $\mathcal{W}_j^{n}$ from its new applicants and its previous waiting list $\mathcal{W}_j^{n-1}$\;
Each BS $j$ keeps the first $m_j$ preferred UEs according to $\mathcal{P}_j^\text{BS}$, and reject the rest of them\;
\For{$k\in\mathcal{R}$}{
$l_k\leftarrow l_k+1$\;
\If{$l_k>J$}{
Remove UE $k$ from $\mathcal{R}$\;
}
}
$n\leftarrow n+1$\;
}
Each BS informs all the UEs in its waiting list about the final association decision\;
\KwOutput{Distributed learning association results $\eta_k^{(t)}$ at UE $k,~k\in\mathcal{K}$}
\caption{(MG-DLB) 
Distributed Action Selection Policy}
\label{MG_DLB_Alg}
\end{algorithm} 

\section{Distributed Multi-Agent Actions Selection Policy for Load Balancing}
\label{MG_DLB_Sec}
\textcolor{black}{
For the distributed approach, we propose a distributed game based on matching theory in which UEs apply to individual BSs and receive a response for association action which ensures load balancing of the responding BS. 
In this distributed approach, no central entity exists and there are no  messages or information exchanges among agents; the user association algorithm operates in a fully distributed manner.
}


\subsection{Distributed Record and Update of Q- and U-tables}
Here there is no central entity which maintains the Q-tables of all UEs. Instead, each UE and each BS stores and maintains its own copy of a single Q-table in a distributed fashion. Specifically, each UE maintains and updates its Q-table and N-table locally. At each learning step $t$, UE $k$ builds its own U-vector ${\mathbf{u}}_k^{(t)}\left(s_k^{(t)} \right)$ based on (\ref{UCB_rule}) and (\ref{NBS}), and sorts the values in this vector in descending order to form its current preference list. 

Each BS $j$ also maintains and updates (based on UEs' reports) its local record of 
a single U-table $\mathbf{U}_j$ composed of $K$ columns, one from each UE corresponding to the action of connecting to this BS $j$. 
Assuming the state space of all agents is of the same size $|\mathcal{S}_1|=...=|\mathcal{S}_k|=...=|\mathcal{S}_K|$, the size of $\mathbf{U}_j$ 
is $|\mathcal{S}_k|\times K$. This assumption is only for simpler notation but is not necessary for the algorithm to work; each BS can maintain $K$ U-vectors,  
each of a specific size $|\mathcal{S}_k|$. 

\begin{figure}
    \centering
    \includegraphics[width=90mm]{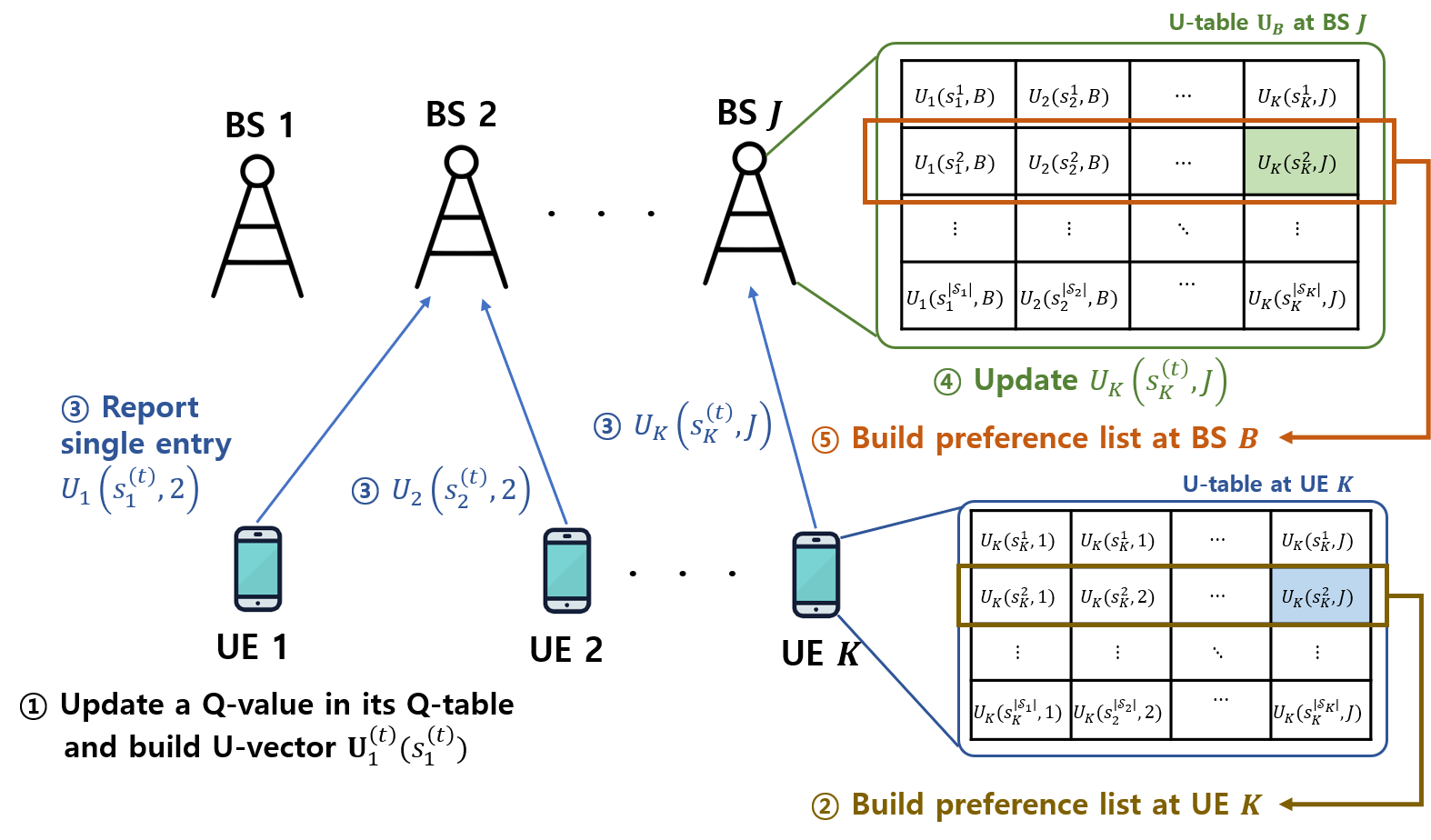}
    \caption{\textcolor{black}{Procedure for U-value update in the proposed distributed action selection policy without CLB.}}
    \label{fig:new_dist}
\end{figure}

At each learning step, each BS $j$ receives up to $m_j$ reports of a new U-value, each from one of its associated UEs, and uses these reports to update its table $\mathbf{U}_j$. 
Then the BS selects the current U-values $\left[{U}_k\left(s_k^{(t)},j \right),~\forall k\in\mathcal{K}\right]$ and sorts them in descending order to  form its preference for that time step. 
\textcolor{black}{The U-value update procedure is illustrated in Fig. \ref{fig:new_dist}.}

\textcolor{black}{
We emphasize that the U-tables maintained at a UE and at a BS are not the same: the U-table at BS $j$ is made up of $K$ different columns, each equals to the $j^\text{th}$ column of the U-table at UE $k,~ (k=1, ..., K)$. Different from the centralized approach, here no entity has all Q-tables or N-tables of all the UEs.
}

\subsection{MG-DLB Load Balancing Action Selection Policy}

We propose a matching game distributed load balancing (MG-DLB) algorithm. At each learning step $t$, the UEs and BSs play a distributed deferred acceptance (DA) matching game \cite{gale1962college} which leads to load-balanced association resulting from each BS controls how many UEs to admit based on its load balancing constraint. 
\textcolor{black}{From the updated U-tables at each UE and each BS, the UEs and the BSs build their own preference list using the U-value vector corresponding to the current state of each UE. Each UE will have a preference list of length $J$, and each BS a preference list of length $K$. All UEs and BSs then participate in the matching game in a distributed fashion, using only the locally stored U-table without the need for a central entity.
Note that the U-table at each BS and each UE is fixed during a game, but will change in the next learning step.
}

\textcolor{black}{
The distributed game works as follows. At each learning step, each UE has a preference list based on its sorted U-vector for its current state $s_k^{(t)}$. Each UE applies first to the BS at the top of its preference list by sending a request to this BS. Each BS also has a preference list based on its sorted U-vector for all UEs' current states $\{s_k^{(t)}\}$. Each BS will wait-list the top $D_j$ requests based on its own preference list, where $D_j$ is its load balancing limit, and reject the rest. Each rejected UE then applies to the next BS on its preference list. The process continues until all UEs have either been wait-listed at a BS or been rejected by all BSs. At that point, each BS commits to admiting all UEs in its waitlist, which provides the association results.
Each game can include up to $J$ application-response rounds between the UEs and BSs. Since the waitlist at BS $j$ only includes up to $D_j$ requests, load balancing at each BS is always guaranteed.
}
A summary of this distributed load balancing is shown in Alg. \ref{MG_DLB_Alg}.

\textbf{\textit{Remark 3:}} 
The MG-DLB is run in a distributed fashion at each UE and each BS and requires no physical host or control entity.  This is in contrast to the WCS-CLB algorithm which is run centrally at the CLB. During an execution of the MG-DLB algorithm, the UEs perform an application-response process with the BSs, where there are at most $J$ application-response rounds. At the end of an MG-DLB execution, each UE receives an association if it is in the final waiting list of a BS, otherwise the UE is unassociated. Each UE then takes the resulting action, and reports the updated U-value for the taken state-action pair to its associated BS. 
All the updating and reporting of U-values for MG-DLB are done in a fully distributed fashion.


\section{Online Multi-agent QL Algorithm for Seamless Handover}
\label{UA_HO_Sec}

In this section, we integrate the above multi-agent decision policies into an \emph{online multi-agent QL algorithm}  that performs user association and handover seamlessly, considering user mobility and handover cost. 

In a dynamic network, user mobility and channel variations can cause UEs to handover from a serving BS to a target BS. The  handover process interrupts data transmission and hence incurs a cost, which is usually translated to a reduction in the data rate. To capture the effect of handover in our proposed algorithm, we apply a handover cost in the learning process.



\begin{algorithm}[t]\footnotesize
\label{Main_Alg}
\SetAlgoLined
\KwInput{Learning rates $\alpha$, discount factor $\gamma$, BSs' quota vector $\mathbf{m}$, randomly generated initial Q-values $Q_k\left(s_k^{(0)},j\right),~\forall k\in\mathcal{K}, s_k\in\mathcal{S}_k, \forall j\in\mathcal{J}$, initialize N-tables $\mathbf{N}_k^{(0)}=\mathbf{0}, \forall k\in\mathcal{K}$\\
\textbf{Initialization}:}
Reporting initial Q-values:\;
~~~ - If centralized: each UE sends its initial Q-table to CLB\;
~~~ - If distributed: each UE sends each column of its initial Q-table to the corresponding BS\;
\textbf{Repeat for} each moving step $n$:\\
\For{each measurement block (MB) $b$}{
\For{$t=1:T$}{
Perform load balancing assignment using WCS-CLB (centralized Agl. \ref{WCS_CLB_Alg}) or MG-DLB (distributed Alg. \ref{MG_DLB_Alg})\;
\textbf{Access and Mobility Management Function (AMF)}:\\
Collects UEs' data rates to calculate network throughput $r(\boldsymbol{\upeta}^{(t)})$ according to (\ref{r_t})\;
\If{$r(\boldsymbol{\upeta}^{(t)})>r(\boldsymbol{\upbeta})$}{
Updates the best-to-date association vector $\boldsymbol{\upbeta}^{(b)}= \boldsymbol{\upeta}^{(t)}$\;
Performs necessary handovers based on $\boldsymbol{\upbeta}^{(b)}$ for data transmission\;}
\textbf{Each UE} $k$:\\
Takes an action $a=\eta^{(t)}_{k}$ which is the next association state\;
Receives reward $R^{(t)}_{k,a}$ and specifies its SINR state from serving BS\;
Updates its Q-value for UA (using (\ref{QL_UR})) or for UA-HO (using (\ref{HO-reward}))\;
Reports $Q_k(s_k^{(t)},a)$ to the CLB (centralized) or to each BS $a$ (distributed)\;
}
\KwOutput{Best-to-date association vector $\boldsymbol{\upbeta}^{(b)}$}
\caption{Online QL for User Association and Handover}
}
\end{algorithm}

We now propose an online QL algorithm which can be implemented in either centralized or distributed fashion by selecting the corresponding load balancing action selection policy as described in Sec. \ref{WCS_CLB_Sec} and Sec. \ref{MG_DLB_Sec}. The proposed algorithm provides the best-to-date association $\boldsymbol{\upbeta}$ at any moving block while the learning process continues indefinitely in the background.


\subsection{Initialization} 
At the beginning of the algorithm, each UE randomly initializes its Q-table. 
For the centralized implementation, each UE sends its entire Q-table to CLB, and the CLB initializes $K$ N-tables as $\mathbf{N}_k=\mathbf{0}, \forall k\in\mathcal{K}$. Since the CLB performs UCB updates and action selections, UEs do not need to keep their N-tables. For the distributed approach, however, each UE $k$ also initializes its N-tables as $\mathbf{N}_k=\mathbf{0}$, then computes its U-table and sends each column of its U-table to the corresponding BS. 

\subsection{Online QL Execution and Updates}
The algorithm starts by selecting a load balancing action selection policy. Centralized or distributed learning can be carried out by employing WCS-CLB (Alg. \ref{WCS_CLB_Alg}) or MG-DLB (Alg. \ref{MG_DLB_Alg}), respectively. 
The online QL algorithm is run continuously after initialization. Each MB $b$ consists of $T$ learning steps, and during each learning step, each UE $k$ performs the following tasks. 

First, it takes an action based on its learning association ($a=\eta_k^{(t)}$). In the centralized implementation, UEs are informed about their learning association by the CLB, but in the distributed implementation each UE knows its learning association as the result of the matching game. 
After taking the action, the UE receives a reward from the selected BS and specifies its SINR state.

Next, the UE updates its Q-value. 
To incorporate a handover cost, we set the reward in (\ref{QL_UR}) by considering a connection-time dependent handover cost. In particular, a handover cost is incurred for a UE if its new learning association differs from the previous learning association (i.e., $\eta_{k}^{(t)}\neq \eta_{k}^{(t-1)}$), resulting in a HO-reward 
$\tilde{R}_{k,a_k^{(t)}}$ obtained as
\begin{align}
  \tilde{R}_{k,a_k^{(t)}}=\left(1-\zeta(\tau_{k}^{(b)})\bar{\delta}(\eta_{k}^{(t-1)},a_k^{(t)})\right)R_{k,a_k^{(t)}}.  
  \label{HO-reward}
\end{align}
Here, $a_k^{(t)}=\eta_{k}^{(t)}$, 
and $\zeta(\tau^{(b)}_k) = C_d e^{-\frac{\tau^{(b)}_k}{10}}+C_0$ represents the handover cost 
where $C_d$ is a soft-cost, $C_0$ is a hard-cost, and $\tau^{(b)}_k$ is the \textit{sojourn time}, the time duration during which the UE has remained associated with its current serving BS up to $t$ \cite{TWC_alireza}, and $R_{k,a_k^{(t)}}$ is the data rate obtained from the connection as in \eqref{R_kvj}. The HO-reward in \eqref{HO-reward} is used in \eqref{QL_UR} for updating Q-values.


\textcolor{black}{
Finally, the UE reports its updated Q-value to the CLB in the centralized implementation, or computes its updated U-value and report this update to the associated BS in the distributed implementation (see Figs. \ref{fig:new_cent} and \ref{fig:new_dist}).
Here, a Q-value (or U-value) is updated by each UE at every learning step, while each moving step of a UE is composed of multiple measurement blocks, each block with $T$ learning steps. 
As such, the proposed algorithm has the ability to adapt to each UE's mobility, which will be clearly verified in the simulation section.
}
This integrated online QL algorithm is summarized in Alg. \ref{Main_Alg}.

\begin{figure}[t]
  \centering
\includegraphics[width=80mm]{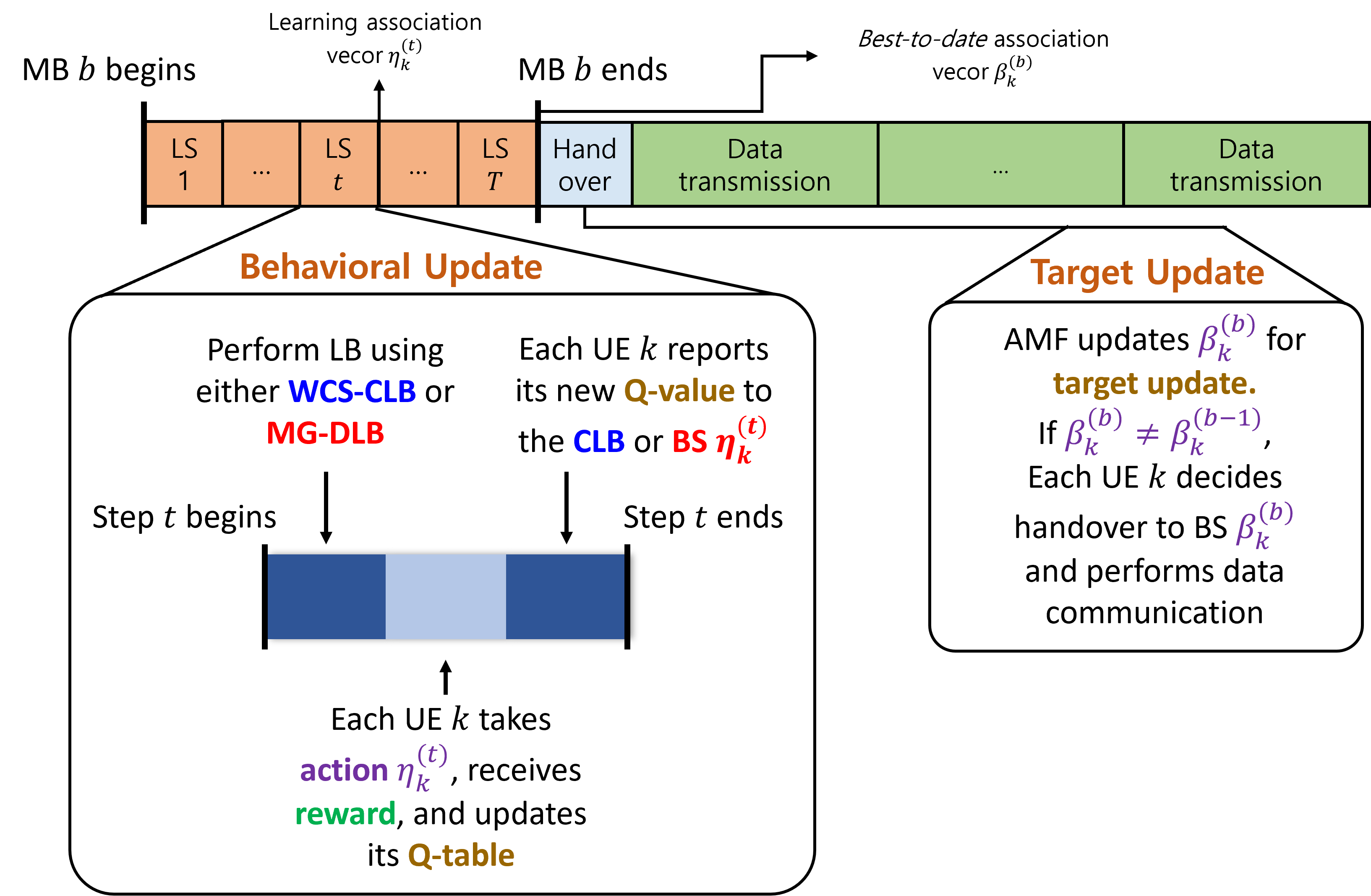}
\caption{Timing diagram for one measurement block (MB) $b$, showing the behavioral (learning) update and the target (association) update.
Each learning step $t$ within MB $b$ runs one inner iteration of the load balancing policy and the updating and reporting process, as described in Alg. \ref{Main_Alg}. After $T$ learning steps, each UE performs handover if $\beta^{(b)}_k\neq \beta^{(b-1)}_k$, followed by data communication between each UE and its associated BS.}
\label{LTS_Block}
\end{figure}


\subsection{Performing Handovers} 
\textcolor{black}{
During the online QL process, 5G core access and mobility management function (AMF)\footnote{\textcolor{black}{The AMF is a central entity in 5G core which receives all connection related information from UEs and is responsible for handover and mobility management tasks \cite{3GPP_5GC}. The functionality of AMF in 5G is similar to mobility management entity (MME) in LTE evolved packet core (EPC).}} maintains the best-to-date association vector $\boldsymbol{\upbeta}^{(b)}$.
At each learning step $t$, the new learning association vector $\boldsymbol{\upeta}^{(t)}$ resulting from the WCS-CLB or MB-DLB algorithm is reported to the AMF. The AMF also collects the user data rates (or SINR values) and calculates the network throughput under the resulting learning association vector.
If there is an improvement compared to the throughput under the best-to-date association, then the AMF updates the best-to-date association vector as $\boldsymbol{\upbeta}^{(b)} = \boldsymbol{\upeta}^{(t)}$.
At the end of MB $b$, each UE $k$ performs handover to BS ${\beta}^{(b)}_k$, and carry out data communication with this new BS.
Note that the vector $\boldsymbol{\upbeta}^{(b)}$ refers to the best association vector only within MB $b$, not of all time. 
Fig. \ref{LTS_Block} summarizes the process of action decision making and association updating in online multi-agent QL in each MB.
}

We note that the AMF is not engaged in the learning process, and is only responsible for collecting all current data rates from the UEs to decide about updating the best-to-date associations for data transmission. 
We thus separate this built-in AMF function from the learning process which can be either centralized (using WCS-CLB) or fully distributed (using MG-DLB). 


\section{Signaling Overhead and Complexity Analysis}
\subsection{Signaling Overhead}
The signaling overhead of the proposed QL algorithm depends on its centralized or distributed implementation. 
For the centralized version of Alg. \ref{Main_Alg}, after performing the load balancing algorithm (WCS-CLB) in each learning step, the CLB informs each UE about its association for learning $\eta_k^{(t)}$ and the AMF 
informs each UE about its association for data transmission
(only if $\beta_k^{(b)}$ changes).
Assuming each signaling of an association variable requires $X_2$ bits, the total number of bits for reporting association results is $KX_2$. 
Then, each UE reports its updated Q-value for the taken (state, action) pair to the CLB. Assuming this reporting requires $X_1$ bits each, the total number of bits sent from all UEs to the CLB is $KX_1$. Thus, the maximum (since $\boldsymbol{\upbeta}$ is not necessarily signaled in every step) signaling overhead per learning step for the centralized load balancer is $X^\text{CLB}=K(X_1+2X_2)$ bits. 
\textcolor{black}{In addition, for updating the best-to-date association, the AFM collects the user data rates or SINR values to compute the sum rate, which requires $X^\text{AMF}=KX_3$ bits, assuming that the data rate information is quantized with $X_3$ bits.}

For the distributed implementation using MG-DLB, 
since the preference lists of UEs and BSs remain unchanged during game, there is no extra signaling other than the one-bit application and response messages which require the maximum (worst case) signaling of $2KJ$ bits. 
If any $\beta_k^{(b)}$ changes, the AMF informs UEs about their associations for data transmission, resulting in $KX_2$ bits.
For reporting, all UEs must report their Q-value to the corresponding BS (instead of reporting them to the CLB).  
Thus, the maximum signaling for this distributed implementation is $X^\text{DLB}=K(X_1+X_2+2J)$ bits.

\textcolor{black}{
In comparison, the centralized, non-learning WCS algorithm in \cite{TWC} requires global network information including instantaneous CSI, which must be collected to a central server. Assuming each element of the channel matrix is quantized with $X_4$ bits, the signaling overhead per measurement for the WCS algorithm is $X^\text{WCS}=KJM_jN_kX_4$. In this non-learning WCS algorithm, each UE must report $M_jN_k$ complex-valued numbers in each MB to each BS. In contrast, the proposed learning algorithm only requires each UE to report a single real U-value (or Q-value) to only one BS (or the CLB) in each learning step, leading to a total of $T$ real-valued numbers reported in each MB. Typically, $T$ is small (around 10 or fewer per MB, as shown later in Sec. VIII.B.1), whereas the product of the numbers of BSs and antennas $JM_jN_k$ can be huge for mmWave communication. 
Thus, the complexity of the non-learning WCS algorithm is significantly higher than either the centralized or distributed QL implementation. The overall comparisons of signaling overhead and complexity are summarized in Table. \ref{table_complexity}.
}

\subsection{Computational Complexity}
Since the Q-value updates in Alg. \ref{Main_Alg} (Lines 14-16) are simple scalar multiplication and addition operations, the computation complexity of the proposed Q-leaning algorithm is dominated by its load balancing algorithm. In what follows, we calculate the cost of each load balancing algorithm separately.

For WCS-CLB (Alg. \ref{WCS_CLB_Alg}), the most computationally consuming steps are Lines 7.
Line 7 involves two operations: computing $K+1$ sum-rewards which is in the order of $\mathcal{O}(K)$, and sorting the resulted sum-rewards which costs $\mathcal{O}(K\log K)$, 
giving the total cost of $\mathcal{O}(K\log K)$ at each \textit{while} loop.
Since the number of iterations of the while loop linearly increases with $K$ \cite{TWC}, the overall cost of Alg. \ref{WCS_CLB_Alg} is of $\mathcal{O}(K^2 \log K)$.

Executing the DA game in MG-DLB (Alg. \ref{MG_DLB_Alg}) incurs a small sorting cost at each player of the game. In Line 2, each UE sorts its U-vector to build its preference list at a cost of $\mathcal{O}(J\log{J})$. 
At each iteration, each BS sorts its applicants to update its waiting list, a procedure which incurs a maximum cost of $\mathcal{O}(K\log{K})$ since the maximum number of applicants for each BS is $K$. The number of iterations in a DA game is at most $J$ \cite{MT}, resulting in the overall complexity of $\mathcal{O}(KJ\log{K})$ at each BS. Since UEs do not perform sorting, the overall complexity at each UE is $\mathcal{O}(J\log{J})$. 
This gives Alg. \ref{MG_DLB_Alg} a total complexity of $J \mathcal{O}(KJ\log K) + K \mathcal{O}(J\log J)$, albeit distributed among all BSs and UEs. 
Since usually $J\ll K$, the computation complexity for MG-DLB is much lower than for WCS-CLB.
 
In short, the signaling overhead of MG-DLB is comparable with that of WCS-CLB, while its computation complexity is much lower than the centralized load balancer and is distributed among all BSs and UEs.

\begin{table}[t]
\caption{Comparison of Signaling Overhead and Complexity}
\begin{center}
\vspace*{-1em}
\begin{tabular}{||c | c | c||} 
 \hline
 \textbf{Scheme} & \makecell{\textbf{Signaling}\\ \textbf{overhead}} & \makecell{\textbf{Computational}\\\textbf{Complexity}} \\ [0.5ex] 
 \hline\hline
 WCS \cite{TWC} & $KJM_jN_kX_3$ & $\mathcal{O}(M_j^2 K^2\log (K))  $ \\ [0.5ex] 
 \hline
 QL-WCS-CLB & $K(X_1+2X_2)$ & $\mathcal{O}(K^2\log K)$  \\[0.5ex] 
 \hline
 QL-MG-DLB & $K(X_1+X_2+2J)$ & $J \mathcal{O}(KJ\log K) + K \mathcal{O}(J\log J)$ \\ [0.5ex] 
 \hline
\end{tabular}
\end{center}
\label{table_complexity}
\vspace*{-2em}
\end{table}


\section{Numerical Results}
\label{Sim_Results}
In this section, we evaluate the performance of our proposed online QL user association and handover algorithm in the downlink of a two-tier HetNet. The MBSs and SBSs operate at 1.8 GHz and 28 GHz, respectively. 
The channels for sub-6 GHz links and mmWave links (composed of 5 clusters and 10 rays per cluster) are generated per MB as described in Sec. \ref{Sys_Ch_models}. To enable 3D beamforming at BSs, each BS is equipped with a UPA of size $8\times 8$ ($M_j=64$).
The number of sub-6 GHz and mmWave antennas at each UE $k$ is $N_k^{\mu\text{W}}=2$ and $N_k^\text{mmW}=4$, respectively. We also  assume that the transmit power of MBS is 10dBm higher than that of SBSs.
The noise power spectral density is $-174$ dBm/Hz, and the respective available bandwidths at each MBS and SBS are 20 MHz and 400 MHz, respectively. 
We consider three network settings for simulations as summarized in Table \ref{Network_settings}. \texttt{Network 2} is the base network used in all simulations. Network nodes are deployed in a $500 \times 500~\textrm{m}^2$ square where the BSs are placed at specific locations and the UEs are distributed randomly according to a homogeneous Poisson point process (PPP).
For learning algorithms implementation, fixed values for learning rate $\alpha=0.9$ and discount factor $\gamma=0.2$ are picked based on extensive evaluations. 

\begin{table}[t]
\caption{Network setting for simulation results}
\begin{center}
\begin{tabular}{||c c c c c||} 
 \hline
 Network & $J_M$ & $J_S$ & $K$ & Quota vector of BSs \\ [0.5ex] 
 \hline\hline
 1 & 1 & 3 & 18 & $\mathbf{m}=[18,6,6,6]$ \\ 
 \hline
 2 & 2 & 4 & 30 & $\mathbf{m}=[18,18,6,6,6,6]$ \\
 \hline
 3 & 2 & 4 & 60 & $\mathbf{m}=[36,36,12,12,12,12]$  \\
 \hline
\end{tabular}
\end{center}
\label{Network_settings}
\vspace*{-2em}
\end{table}

Throughout this section, the proposed QL algorithms are compared with the following benchmarks:
\begin{itemize}[leftmargin=*]
    \item \textbf{WCS} \cite{TWC}: Centralized optimization (non-learning) algorithm for load-balancing association achieving near optimal network sum rate.
    WCS must collect global network CSI and be re-run for each measurement update per MB (see Table I), hence is used as an upper bound for network sum rate performance.

    \item \textbf{3GPP Max-SINR} \cite{3GPP_5GC}: Each UE coonects to the BS providing the highest max-SINR, and we drop users if a BS becomes overloaded to ensure load balancing.
\end{itemize}


\begin{figure}[t]
\centering
\subfloat[Total expected rewards of all UES (upper) and expected reward of two typical UEs (lower) for the proposed centralized QL algorithm (\texttt{QL-WCS-CLB})]{\includegraphics[width=85mm]{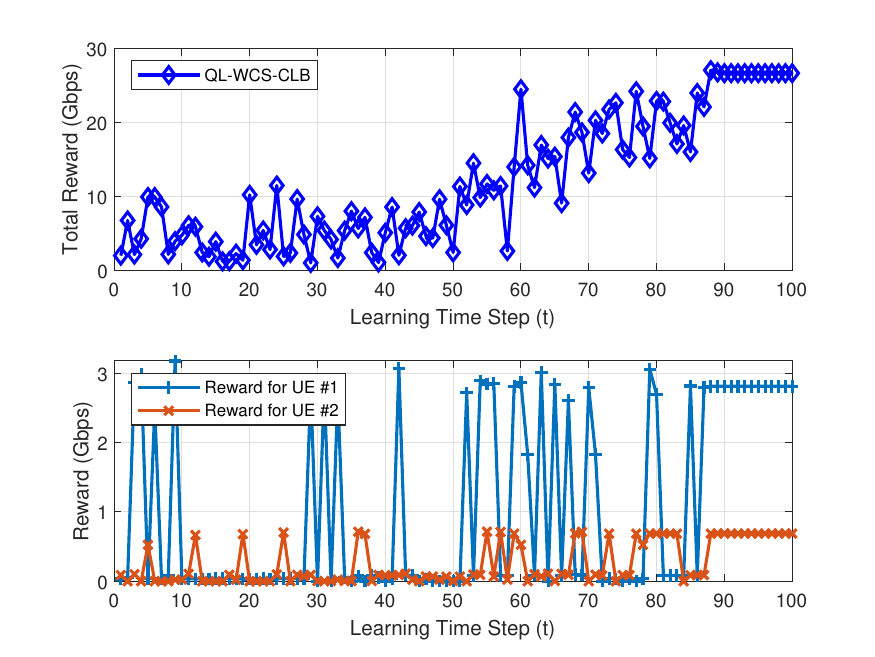}}\\
\subfloat[The average user data rate of the proposed centralized (\texttt{QL-WCS-CLB}) and distributed (\texttt{QL-MG-DLB}) QL algorithms]{\includegraphics[width=85mm]{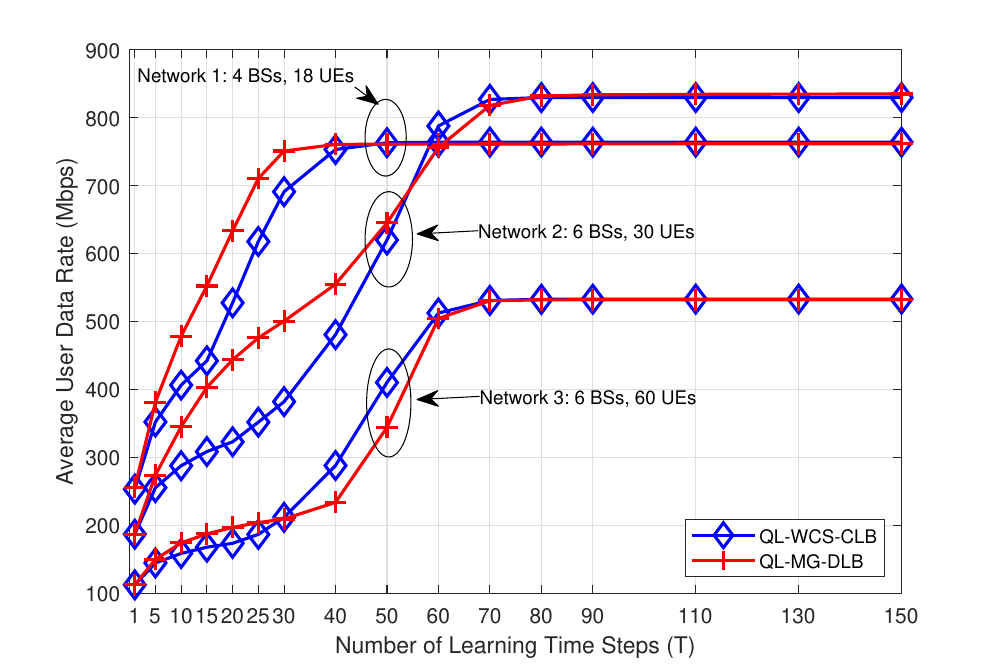}}
\caption{Convergence behavior of the proposed algorithms in static networks (no UE mobility).}
\label{QL_Reward_SR_vs_time}
\end{figure}
\subsection{Learning Performance in Static Networks}
\label{Learn_analysis_Sim}
First we consider static networks in which users are stationary with random but fixed locations (no mobility). 
We implemented both the centralized and distributed versions of the proposed QL algorithm: i) \texttt{QL-WCS-CLB} which employs the WCS-CLB algorithm for load balancing (Alg. \ref{WCS_CLB_Alg}), and ii) \texttt{QL-MG-DLB} which performs the MG-DLB algorithm for distributed load balancing (Alg. \ref{MG_DLB_Alg}). 

\subsubsection{Learning convergence and achievable data rate}
Fig. \ref{QL_Reward_SR_vs_time}.a shows convergence plots of the total expected reward of all UEs (upper figure) and the expected rewards of two typical UEs (lower figure) with respect to learning steps for \texttt{QL-WCS-CLB} in \texttt{Network 2}. 
It is clear that the total reward improves as learning step grows. The results show that the algorithm converges in a reasonable number of learning steps (around 86 in this setting).

Fig. \ref{QL_Reward_SR_vs_time}.b depicts the best-to-date average per-user data rate versus the number of learning steps for the centralized and distributed QL algorithms for the networks defined in Table \ref{Network_settings}.
Interestingly, for all network sizes, the distributed \texttt{QL-MG-DLB} performs better than the centralized \texttt{QL-WCS-CLB} for a low number of learning steps.
The convergence for both the centralized and distributed QL algorithms is similar and quite fast; for example, in \texttt{Network 1}, both algorithms reach the maximum performance after only 40 steps. Also, it is worth noting that the average per-user data rate in \texttt{Network 3} is lower than \texttt{Network 2}, while both the number of UEs and quota of BSs are doubled. This is due to the fact that, the increased number of UEs in \texttt{Network 3} resulted in higher interference in the network, and hence lower per-user data rate. The sum rate among all users or the network throughput, however, is the highest for \texttt{Network 3}.

\textcolor{black}{
These results show that our multi-agent QL algorithms are well converged even though the agents do not share any information, nor communicate, nor have access to common information. 
The convergence of multi-agent QL when only local information is exploited at each agent has, in generally, not been proven theoretically, but only shown via empirical means \cite{MA_proof2}. In the same manner, here we show the convergence of our proposed algorithms via simulation.
Due to the joint action decision at the CLB in \texttt{QL-WCS-CLB} or at the BSs in \texttt{QL-MG-DLB}, convergence of the proposed algorithms is guaranteed
even though no common state information is available to the agents.
}


\begin{figure}
  \centering
  \includegraphics[width=90mm]{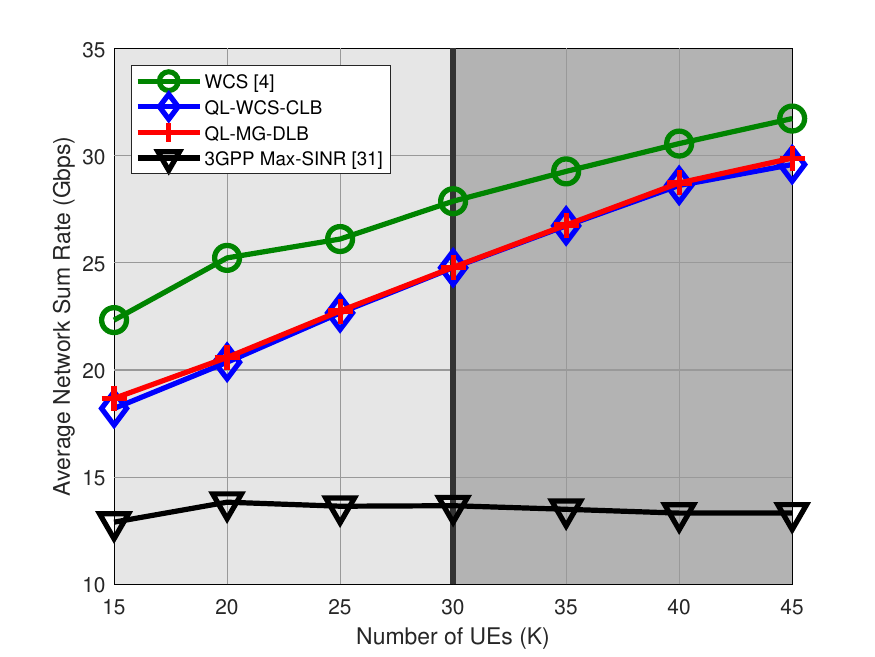} 
  \vspace*{-1em}
  \caption{Effect of different loading scenarios on the average network throughput in a static \texttt{Network 2} with $T=100$ (no UE mobility). The transmit powers at MBSs and SBSs are 45 dBm and 35 dBm.}
  \label{SR_vs_Load}
\vspace*{-1em}
\end{figure}


\subsubsection{Learning adaption to network load}
Fig. \ref{SR_vs_Load} evaluates throughput performance of the proposed QL algorithms under different network loading scenarios for \texttt{Network 2}: 1) underload ($K<30$), 2) critical load ($K=30$), and 3) overload ($K>30$). 

Both the centralized and distributed QL algorithms show similar performance, and outperform \texttt{3GPP Max-SINR} algorithm by between 48-130\%.
They also reach closely the performance of \texttt{WCS} algorithm, with a gap of about 9\% under light load scenario (15 UEs) and 4\% under heavy load scenario (45 UEs).
For all learning-based algorithms, we observe that the network throughput increases with more users, even under overload scenarios, while \texttt{3GPP Max-SINR} exhibits an almost flat network throughput which is slightly decreasing when overloading. This result illustrates the robustness and effectiveness of the proposed learning algorithms to associate users even under overload scenarios to keep the throughput increase while satisfying load balancing.

\begin{figure}
  \centering
  \includegraphics[width=80mm]{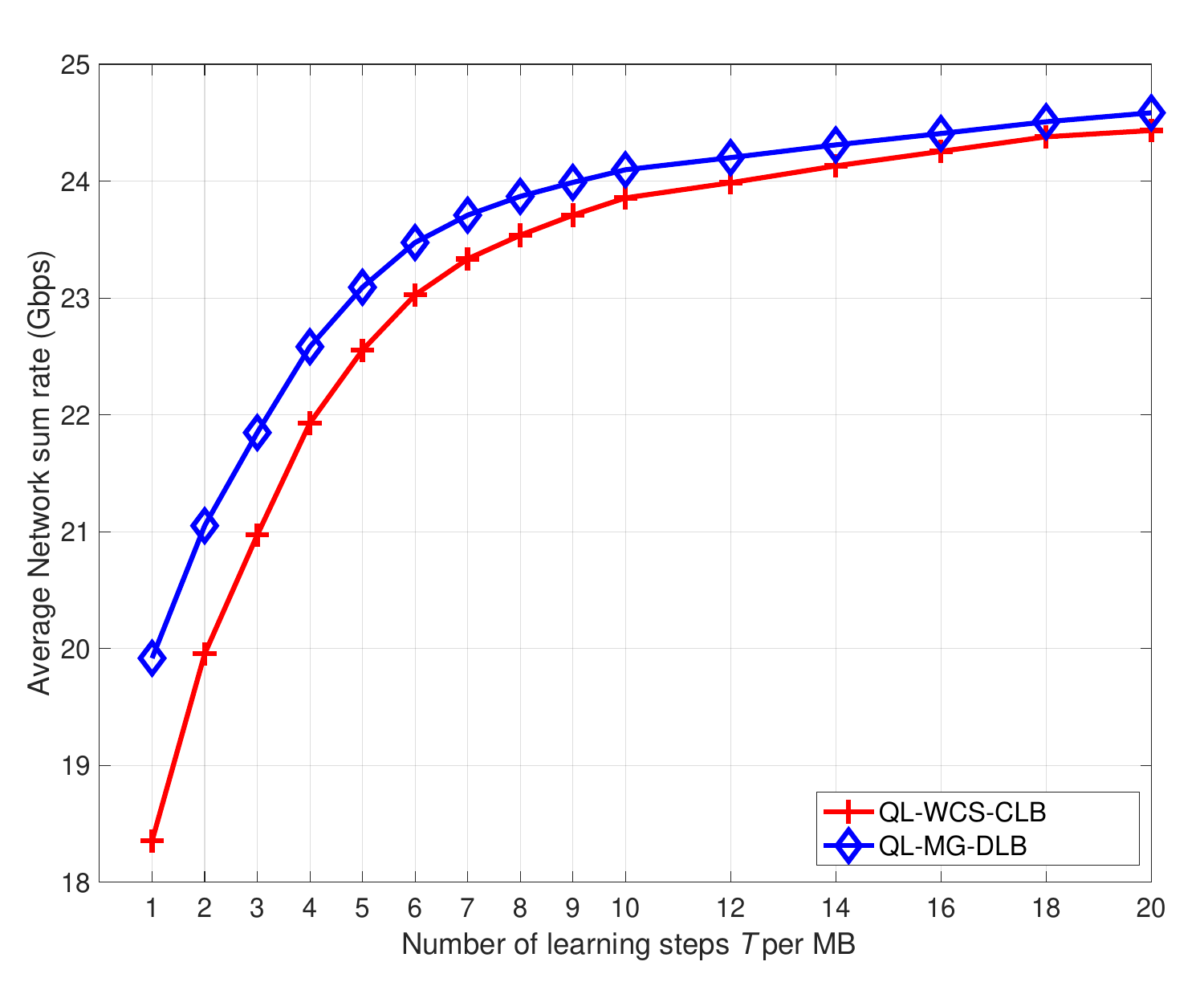}
  \vspace*{-1em}
  \caption{Effect of the number of learning steps $T$ per MB, on the network throughput in dynamic \texttt{Network 2} with UE mobility. In each moving step, 30\% of UEs are moving, each with a random velocity $V_{k,n}\in \{1, ..., 10\}$ m/s. The data rate values are captured after 5 moving steps, when the throughput starts to converge to a stable value.}
  \label{SR_vs_NumMeasPerMB}
\vspace*{-1em}
\end{figure}

\subsection{Handover in Dynamic Networks}

Next we evaluate the performance of the proposed online QL algorithm, Alg. \ref{Main_Alg}, in dynamic networks where UEs are mobile and can experience handover. For these simulations, we focus on \texttt{Network 2}, and assume that the duration of each MB is $t^\text{MB}=480$ ms, and in each moving step, 30\% of UEs are moving.

\subsubsection{Number of learning steps per block}
Fig. \ref{SR_vs_NumMeasPerMB} evaluates the effect of number of learning steps $T$ per MB on the average network throughput after $n=5$ moving steps, when the network throughput starts to converge to a stable value. 
It can be observed that the distributed QL algorithm slightly performs better especially at low values of $T$.
The results show that at small values of $T$, increasing the number of learning steps $T$ per block has a significant impact on increasing the network throughput, as more steps provides the learning algorithms with more data for real-time updates. This increase in throughput, however, diminishes as $T$ increases beyond around 10 steps per block. For practical purposes, $T$ should be kept relatively small to reduce system overhead. Results in Fig. \ref{SR_vs_NumMeasPerMB} suggests that $T=6$ is a reasonable value to achieve good throughput performance, thus we use $T=6$ in each MB in all subsequent simulations.

\subsubsection{Mobility trajectories and handovers example}

\begin{figure}
    \centering
    \includegraphics[width=95mm]{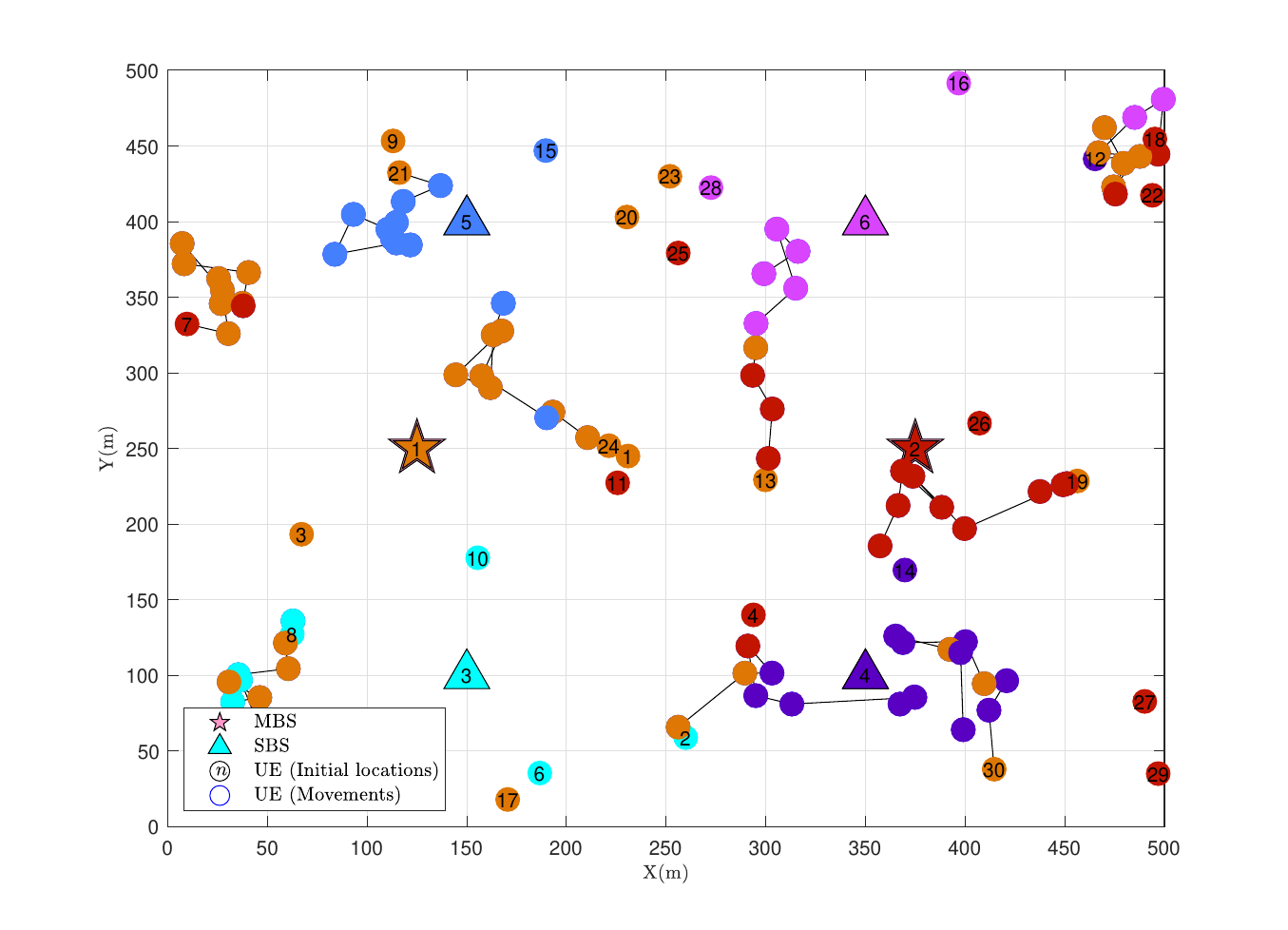}
    \vspace*{-2em}
    \caption{A layout example of dynamic \texttt{Network 2} where each UE is labeled with a number at its initial location. In this setting, 9 UEs (30\% of UEs) move randomly at each moving step at $V=6$~{km/h} (walking). Each small circle represents the location of a UE at a particular way point, and the color of the circle corresponds to the color of the associated BS at that way point. A change of color along a UE's trajectory signifies a handover.}
    \label{fig:traj}
\end{figure}

Under mobility at walking speed ($V=6~\text{km/h}$)  with random initial UE locations, 
Fig. \ref{fig:traj} shows the UE trajectories and associations of the UEs along their movement trajectories. 
The figure shows multiple instances where specific users switch their associations along their trajectory, hence handovers occur. It can also be observed that the associated BS of a UE is not always the closest BS. Take UE number 13 for example, during its movements, the UE switches associations from the orange BS (initial BS) to red then to red and then to pink, in order to satisfy load balancing. Note that since the proposed online QL algorithms takes into account the handover cost as in \eqref{HO-reward}, the average handover frequency is in fact kept to a minimum, as will be verified in Fig. \ref{SR_HoR_vs_MS}.


\subsubsection{Real-time network throughput at different mobility speeds}
Fig. 8 compares the network throughput of the proposed online multi-agent QL algorithm in a static network and dynamic networks with different UE mobility categories: 1) walking ($V=6~\text{km/h}$), 2) biking ($V=17~\text{km/h}$), and 3) suburban driving ($V=40~\text{km/h}$). All dynamic networks start from an initial state of complete rest with random initial load-balanced associations. The figure shows network throughputs achieved by online learning Alg. \ref{Main_Alg} in real time, averaged over different sets of initial locations of UEs.

\textcolor{black}{
The proposed algorithms achieve a network throughput in dynamic settings close to that in the static setting (at 94\% throughput for $V=6~\text{km/h}$). 
The figure also shows that faster user mobility speed requires more moving steps for convergence. 
The initial ramping up during the first few moving steps is due to a limited number of measurements and short history at the beginning, and as the number of MBs increases, network throughput performance improves because the learning process has accumulated a longer and richer history.
A higher mobility speed incurs a more dynamic Q-table (or U-table), but 
the proposed online QL algorithm can adapt to these dynamics by updating Q-values (or U-values) at every learning step, converging to a stable performance with accumulated learning history.
While there is a decrease in network throughput with the increase in user mobility speed as expected, the online QL algorithm is able to converge and reach a stable throughput at each mobility speed after at most $4$ moving steps.
}


\begin{figure}
    \centering
    \includegraphics[width=90mm]{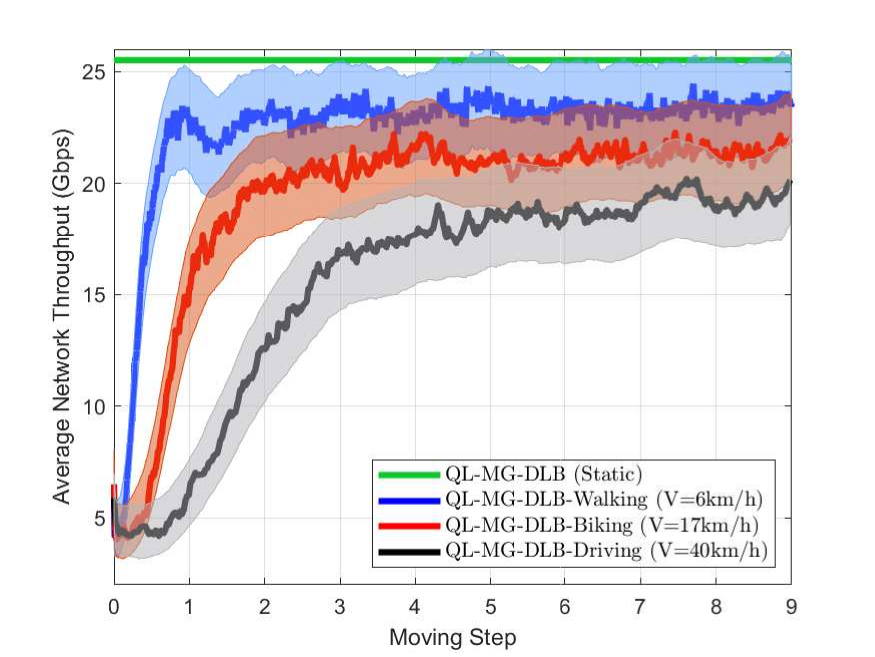}
    \caption{Effect of mobility speed on the convergence of network throughput in dynamic \texttt{Network 2}, where shaded areas indicate standard deviations. For each network at a particular speed setting, 30\% of UEs are moving at any one time.}
    \label{fig:new}
    \vspace*{-1em}
\end{figure}

\begin{figure}[t]
\centering
\includegraphics[width=80mm]{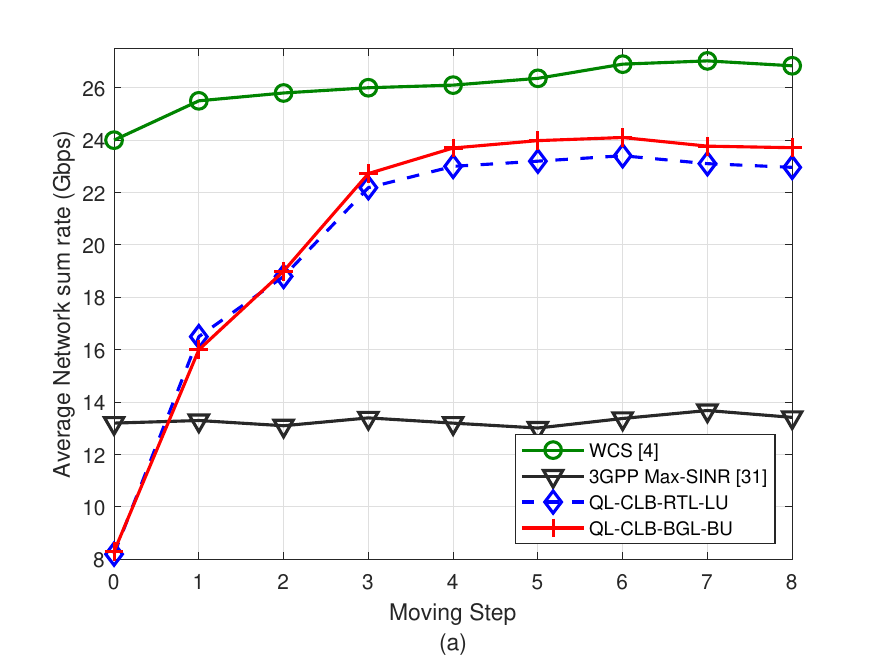}
\hspace*{.5em}
\includegraphics[width=80mm]{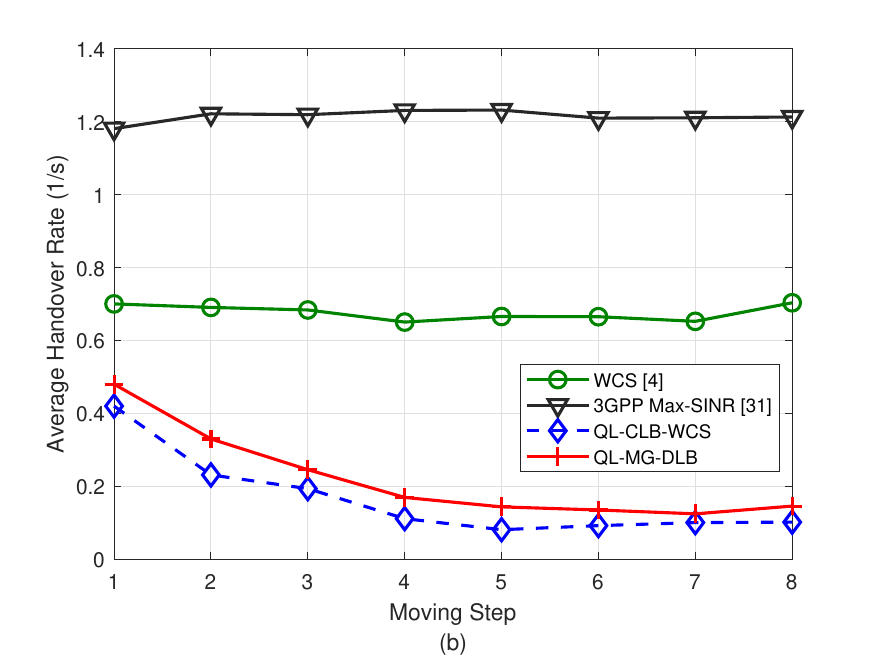}
\vspace*{-1em}
\caption{Comparisons of a) the average network throughput rate, and b) the average handover rate of the proposed online QL Alg. \ref{Main_Alg} in constrast with \texttt{WCS} \cite{TWC} and \texttt{3GPP Max-SINR} algorithms in a dynamic \texttt{Network 2} where 30\% of UEs are moving at each moving step with a random velocity $V_{k,n}\in \{1, ..., 10\}$ m/s.}
\label{SR_HoR_vs_MS}
\vspace*{-1em}
\end{figure}

\subsubsection{Throughput and handover frequency comparison} 
Fig. \ref{SR_HoR_vs_MS} depicts the average network throughput and average handover rate of several user association schemes versus moving steps. 
For this simulation, we implemented the distributed algorithm \texttt{QL-MG-DLB} with the reward function in \eqref{HO-reward}, while for the centralized version \texttt{QL-WCS-CLB} we replaced $\eta_k^{(t-1)}$ in \eqref{HO-reward} with $\beta_k^{(b-1)}$ as in \cite{TWC_alireza} as it provides the lowest handover rate based on extensive simulations. 

The proposed QL algorithms significantly outperform the \texttt{3GPP Max-SINR} scheme in both network throughput and handover rate. 
Fig. \ref{SR_HoR_vs_MS}.a shows that the centralized and distributed QL algorithms achieve 
a network throughput 
within 87\% and 89\% respectively 
of the benchmark \texttt{WCS}, while almost doubling the throughput achieved by 3GPP. 
Fig. \ref{SR_HoR_vs_MS}.b 
shows that the proposed learning algorithms achieve significantly lower handover rates than both \texttt{WCS} and \texttt{3GPP Max-SINR}, at 
around an order of magnitude lower than that of 3GPP.

Thus compared to 3GPP, the proposed QL algorithm reduces frequent handover by a factor of $10$ while almost doubling the achievable throughput in mmWave networks. 
These results point to a high level of real-time adaptability and effectiveness of the proposed QL algorithms with handover under user mobility.
Furthermore, the proposed QL algorithms achieve better handover frequency than WCS at a lower network sum rate. This presents an interesting trade-off between network sum rate and handover frequency.



%

\ifCLASSOPTIONcaptionsoff
  \newpage
\fi




\begin{thebibliography}{10}
\providecommand{\url}[1]{#1}
\csname url@samestyle\endcsname
\providecommand{\newblock}{\relax}
\providecommand{\bibinfo}[2]{#2}
\providecommand{\BIBentrySTDinterwordspacing}{\spaceskip=0pt\relax}
\providecommand{\BIBentryALTinterwordstretchfactor}{4}
\providecommand{\BIBentryALTinterwordspacing}{\spaceskip=\fontdimen2\font plus
\BIBentryALTinterwordstretchfactor\fontdimen3\font minus
  \fontdimen4\font\relax}
\providecommand{\BIBforeignlanguage}[2]{{%
\expandafter\ifx\csname l@#1\endcsname\relax
\typeout{** WARNING: IEEEtran.bst: No hyphenation pattern has been}%
\typeout{** loaded for the language `#1'. Using the pattern for}%
\typeout{** the default language instead.}%
\else
\language=\csname l@#1\endcsname
\fi
#2}}
\providecommand{\BIBdecl}{\relax}
\BIBdecl

\bibitem{Andrews}
Q.~Ye~et al., ``User association for load balancing in heterogeneous cellular
  networks,'' \emph{IEEE Trans. Wireless Commun.}, vol.~12, no.~6, pp.
  2706--2716, 2013.

\bibitem{Caire}
D.~Bethanabhotla, O.~Y. Bursalioglu, H.~C. Papadopoulos, and G.~Caire,
  ``{Optimal user-cell association for massive MIMO wireless networks},''
  \emph{IEEE Trans. Wireless Commun.}, vol.~15, no.~3, pp. 1835--1850, 2016.

\bibitem{TVT_irs}
E.~M. Taghavi, R.~Hashemi, A.~Alizadeh, N.~Rajatheva, M.~Vu, and M.~Latva-aho,
  ``Joint active-passive beamforming and user association in irs-assisted
  mmwave cellular networks,'' \emph{IEEE Trans. Veh. Technol.}, pp. 1--14,
  2023.

\bibitem{TWC}
A.~Alizadeh and M.~Vu, ``Load balancing user association in millimeter wave
  mimo networks,'' \emph{IEEE Trans. Wireless Commun.}, vol.~18, no.~6, pp.
  2932--2945, 2019.

\bibitem{LimitedCSI20}
P.~{Han}, Z.~{Zhou}, and Z.~{Wang}, ``User association for load balance in
  heterogeneous networks with limited csi feedback,'' \emph{IEEE Commun.
  Lett.}, vol.~24, no.~5, pp. 1095--1099, 2020.

\bibitem{Khosravi20}
S.~Khosravi, H.~S. Ghadikolaei, and M.~Petrova, ``Learning-based load balancing
  handover in mobile millimeter wave networks,'' in \emph{Proc. IEEE Globecom},
  Dec 2020, pp. 1--7.

\bibitem{mnih2015human}
V.~Mnih~et al., ``Human-level control through deep reinforcement learning,''
  \emph{nature}, vol. 518, no. 7540, pp. 529--533, 2015.


\bibitem{ML_Intro_Book}
R.~S. {Sutton} and A.~G. {Barto}, \emph{{Reinforcement Learning: An
  Introduction}}.\hskip 1em plus 0.5em minus 0.4em\relax The MIT Press, 2015.

\bibitem{QL_UA19}
D.~Li~et al., ``{User Association and Power Allocation Based on Q-Learning in
  Ultra Dense Heterogeneous Networks},'' in \emph{Proc. IEEE GLOBECOM}, Dec.
  2019.

\bibitem{DRL_UA19}
N.~Zhao~et al., ``Deep reinforcement learning for user association and resource
  allocation in heterogeneous cellular networks,'' \emph{IEEE Trans. Wireless
  Commun.}, vol.~18, no.~11, pp. 5141--5152, 2019.

\bibitem{DRL_assoc1}
Q.~Zhang, Y.-C. Liang, and H.~V. Poor, ``Intelligent user association for
  symbiotic radio networks using deep reinforcement learning,'' \emph{IEEE
  Trans. Wireless Commun.}, vol.~19, no.~7, pp. 4535--4548, Jul 2020.

\bibitem{Sana19}
M.~Sana, A.~De~Domenico, W.~Yu, Y.~Lostanlen, and E.~C.~Strinati, ``Multi-agent
  reinforcement learning for adaptive user association in dynamic mmwave
  networks,'' \emph{IEEE Trans. Wireless Commun.}, vol.~19, no.~10, pp.
  6520--6534, Oct 2020.

\bibitem{5GHandover}
V.~Yajnanarayana, H.~Rydén, and L.~Hévizi, ``5{G} handover using
  reinforcement learning,'' in \emph{Proc. IEEE 3rd 5G World Forum (5GWF)},
  2020, pp. 349--354.

\bibitem{new_ref4}
K.~Tan, D.~Bremner, J.~Le~Kernec, Y.~Sambo, L.~Zhang, and M.~A. Imran,
  ``Intelligent handover algorithm for vehicle-to-network communications with
  double-deep {Q}-learning,'' \emph{IEEE Trans. Veh. Technol.}, vol.~71, no.~7,
  pp. 7848--7862, July 2022.

\bibitem{new_ref3}
C.~Lee, J.~Jung, and J.-M. Chung, ``Intelligent dual active protocol stack
  handover based on double {DQN} deep reinforcement learning for 5{G} mmwave
  networks,'' \emph{IEEE Trans. Veh. Technol.}, vol.~71, no.~7, pp. 7572--7584,
  Jul 2022.

\bibitem{new_ref1}
J.~Moon, S.~Kim, H.~Ju, and B.~Shim, ``Energy-efficient user association in
  mmwave/thz ultra-dense network via multi-agent deep reinforcement learning,''
  \emph{IEEE Trans. Green Commun. Netw.}, vol.~7, no.~2, pp. 692--706, Jun
  2023.

\bibitem{new_ref2}
H.-H. Chang, H.~Chen, J.~Zhang, and L.~Liu, ``Decentralized deep reinforcement
  learning meets mobility load balancing,'' \emph{IEEE/ACM Trans. Networking},
  vol.~31, no.~2, pp. 473--484, Apr 2023.

\bibitem{TWC_alireza}
A.~Alizadeh and M.~Vu, ``Reinforcement learning for user association and
  handover in mmwave-enabled networks,'' \emph{IEEE Trans. Wireless
  Communications}, vol.~21, no.~11, pp. 9712--9728, Nov 2022.

\bibitem{wcl_lim}
B.~Lim and M.~Vu, ``Distributed multi-agent deep {Q}-learning for load
  balancing user association in dense networks,'' \emph{IEEE Wireless Commun.
  Lett.}, vol.~12, no.~7, pp. 1120--1124, Jul 2023.

\bibitem{3GPP901}
{3rd Generation Partnership Project (3GPP)}, ``{Study on channel model for
  frequencies from 0.5 to 100 GHz},'' Technical Report 38.901, Jun. 2018, v.
  15.0.0.

\bibitem{Nokia}
T.~A. Thomas, H.~C. Nguyen, G.~R. MacCartney, and T.~S. Rappaport, ``{3D mmWave
  channel model proposal},'' in \emph{Proc. IEEE 80th Veh. Technol. Conf. (VTC
  Fall)}, 2014, pp. 1--6.

\bibitem{MRWP_Mobility}
X.~Lin, R.~Krishna~Ganti, P.~J. Fleming, and J.~G. Andrews, ``Towards
  understanding the fundamentals of mobility in cellular networks,'' \emph{IEEE
  Trans. Wireless Commun.}, vol.~12, no.~4, pp. 1686--1698, 2013.

\bibitem{3GPP331}
{3rd Generation Partnership Project (3GPP)}, ``{5G NR Radio Resource Control
  (RRC); Protocol specification},'' Technical Specification 38.331, May. 2022,
  v. 17.0.0.

\bibitem{sukhbaatar2016learning}
S.~Sukhbaatar, A.~Szlam, and R.~Fergus, ``Learning multiagent communication
  with backpropagation,'' \emph{NIPS}, vol.~29, 2016.

\bibitem{singh2018learning}
A.~Singh, T.~Jain, and S.~Sukhbaatar, ``Learning when to communicate at scale
  in multiagent cooperative and competitive tasks,'' \emph{arXiv preprint
  arXiv:1812.09755}, 2018.

\bibitem{lowe2017multi}
R.~Lowe, Y.~I. Wu, A.~Tamar, J.~Harb, O.~Pieter~Abbeel, and I.~Mordatch,
  ``Multi-agent actor-critic for mixed cooperative-competitive environments,''
  \emph{Advances in neural information processing systems}, vol.~30, 2017.

\bibitem{iqbal2019actor}
S.~Iqbal and F.~Sha, ``Actor-attention-critic for multi-agent reinforcement
  learning,'' in \emph{International conference on machine learning}.\hskip 1em
  plus 0.5em minus 0.4em\relax PMLR, 2019, pp. 2961--2970.


\bibitem{sutton2018reinforcement}
R.~S. Sutton and A.~G. Barto, \emph{Reinforcement learning: An
  introduction}.\hskip 1em plus 0.5em minus 0.4em\relax MIT press, 2018.

\bibitem{van2016deep}
H.~Van~Hasselt, A.~Guez, and D.~Silver, ``Deep reinforcement learning with
  double {Q}-learning,'' in \emph{Proceedings of the AAAI conference on
  artificial intelligence}, vol.~30, no.~1, 2016.

\bibitem{gale1962college}
D.~Gale and L.~S. Shapley, ``College admissions and the stability of
  marriage,'' \emph{The American Mathematical Monthly}, vol.~69, no.~1, pp.
  9--15, 1962.

\bibitem{3GPP_5GC}
3GPP, ``{System Architecture for the 5G System},'' {3rd Generation Partnership
  Project (3GPP)}, Technical Specification 23.501, Jun. 2020, version 15.12.0.

\bibitem{MT}
A.~{Alizadeh} and M.~{Vu}, ``Distributed user association in {B5G} networks
  using early acceptance matching games,'' \emph{IEEE Trans. Wireless Commun.},
  vol.~20, no.~4, pp. 2428--2441, Apr. 2021.

\bibitem{MA_proof2}
M.~Tan, ``Multi-agent reinforcement learning: Independent vs. cooperative
  agents,'' in \emph{Proceedings of the tenth international conference on
  machine learning}, 1993, pp. 330--337.

\end{thebibliography}
%

\section{Conclusion}
We proposed online multi-agent QL algorithms for user association and handover in dense mmWave networks, where each user acts as an agent employing either a centralized or a distributed load balancing policy to ensure all BSs load balancing. In the centralized approach, the action selection for UEs is done by a CLB, while in the distributed approach, each UE takes an action based on its local information via playing a distributed matching game. We apply a mobility and measurement model and integrate a handover cost in the reward for users. Numerical results show significant network throughput and handover rate improvements over the conventional 3GPP max-SINR scheme, where the throughput reaches close to that of the benchmark WCS algorithm at a significantly lower handover rate. 
Both the proposed centralized and distributed algorithms featured low-complexity and fast convergence, making them suitable for real-time operation in highly dynamic networks.

\bibliographystyle{IEEEtran}

\end{document}